\documentclass[aps,twocolumn]{revtex4-1}
\usepackage{amsmath,amssymb}
\usepackage{slashed}
\usepackage{graphicx}
\usepackage{bm}
\usepackage[T1]{fontenc}
\usepackage[utf8]{inputenc}
\usepackage{gauss} 
\usepackage[top=1.0in,bottom=1.0in,left=1.0in,right=1.0in]{geometry}
\usepackage[colorlinks,linkcolor=blue,citecolor=blue]{hyperref}
\newcommand{\be}{\begin{equation}}
\newcommand{\ee}{\end{equation}}
\newcommand{\bea}{\begin{eqnarray}}
\newcommand{\eea}{\end{eqnarray}}
\newcommand{\ba}{\begin{eqnarray}}
\newcommand{\ea}{\end{eqnarray}}

\begin{document}

\title{Hadronic structure on the light-front IX .\\
Orbital-spin-isospin wave functions of baryons}

\author{Nicholas Miesch}
\email{nicholas.miesch@stonybrook.edu}
\affiliation{Center for Nuclear Theory, Department of Physics and Astronomy, Stony Brook University, Stony Brook, New York 11794--3800, USA}

\author{Edward Shuryak}
\email{edward.shuryak@stonybrook.edu}
\affiliation{Center for Nuclear Theory, Department of Physics and Astronomy, Stony Brook University, Stony Brook, New 
York 11794--3800, USA}

\author{Ismail Zahed}
\email{ismail.zahed@stonybrook.edu}
\affiliation{Center for Nuclear Theory, Department of Physics and Astronomy, Stony Brook University, Stony Brook, New York 11794--3800, USA}

\begin{abstract} This paper which is part of a series,  is devoted to several technical issues. In the first part of the paper, we discuss 
the usual wavefunctions in the CM frame for baryons, by clarifying  the representations of the three-quark permutation group $S_3$. We extend the analysis for up to five ``spinors"
with $\rho,\lambda$-symmetry, and 
derive explicitly the totally symmetric wavefunctions modulo color.
They are explicitly used to describe the excited nucleons $N^*$ states, in the P- and D-shell. We also show how to use symbolic operations  in Mathematica, in
spin-tensor notations to make explicit these states. For the S- and P-shells, the totally antisymmetric wavefunctions are given, and the pertinent matrix elements for the spin-dependent operators calculated,
including the mixing between states with different total spin $S$. In the second part of the paper we 
turn to the light front wavefunctions, with an emphasis  on the longitudinal wavefunctions, 
with a novel basis set.
We also discuss their symmetries under permutations,  and select the proper combinations for the transverse and longitudinal excitations for $N^*$ on the light front.
 \end{abstract}

\maketitle

\section{Introduction}

%Let us start with brief account of the historic development.
The nonrelativistic quark model, which
originated in the 1960's, was first based on spin-flavor $SU(6)$ symmetry. Using its multiplets it was possible to
define  the baryon wavefunctions for the lowest S-shell,  and calculate their quantum numbers and important properties, e.g. magnetic moments.
With  the advent of QCD and quarkonia in the  1970's, perturbative and  confining
forces between quarks were added, including spin-dependent interactions, leading to a
qualitatively successful spectroscopy. One important simplification was the use of
 quadratic (oscillatory) confinement, see e.g. \cite{Isgur:1978wd,Isgur:1979be}.

 The extension of this theory to  excited baryons, especially to the
 P-shell with negative parity~\cite{Isgur:1978xj}, revealed a number of issues. To simplify 
some of the symmetry issues 
for three-light-quark 
states $qqq$, they focused on the mixed $qqs$ states ($\Sigma,\Lambda$) ones, and then discussed the light quark limit $m_s\rightarrow m_{u,d}$.
 
 They have shown  that the perturbative predictions from the one-gluon exchange worked qualitatively well only for spin-spin and tensor forces, while the spin-orbit calculated in the same approximation was absent. A key shortcoming at the time, was the missing $D$-shell states predicted
 but not observed
(``missing resonances problem"). Further refinements,
including in particular the inclusion of relativized quarks, were made by Capstick and Isgur \cite{Capstick:1985xss}. 
With time, the issue of ``missing resonances" faded away, as the states in the second and the third shells were nearly all observed. 
 
More recent quarks models for mesons and baryons were defined using light front variables, with Light Front Wave Functions (LFWFs). The advantage
of the light front, 
is the potential to relate the LFWFs to partonic observables, DAs, PDFs, GPDs etc. This formulation, treats democratically heavy and light quarks,
with no need for a  nonrelativitic approximation. The LFWFs for
hadrons made of light and heavy quarks are developed in the same setting, which is very convenient for the discussion of the multi-quark hadrons discovered in the last decade. 

The first step for the description of the LFWFs, 
 has been made by Ji et al \cite{Ji:2002xn},  who
have classified the components of the spin-up proton in terms of spin/orbital helicities. In this formulation the wavefunctions of the 3-quark sector are multicomponent, e.g. the (ground shell) nucleon is ascribed to possess six unrelated functions. For excited baryon states the issues related with 
required quark permutation symmetry were not resolved. Also, the LFWFs for baryonic excitations with  nontrivial orbital-spin-isospin wave functions, has not been addressed to our knowledge.
%were not really considered.

The aims of this technical paper are twofold. First, we wish to clarify how to
build  symmetric representations of quark permutation symmetry $S_3$, as it follows from
quark Fermi statistics. 
 There is no need to take the indirect root via strange baryons
 again and again,  we re-derive
the explicit wavefunctions for the ground S-shell, negative parity P-shell and D-shell nucleons. We propose to use  standard  spin-tensor
forms with many components, conveniently manipulated symbolically in Mathematica.
We then turn to the derivation of the LFWFs, clarifying their orbital-spin-isospin structure.

 Our other aim is to use the available phenomenological information on nucleon excitations,  and translate it 
 into matrix elements of the central and spin-dependent interquark forces. Since the 1970's, spectroscopists still use the one-gluon exchange 
 forces plus linear potential due to the electric flux tubes. But even for heavy quarkonia, we found in \cite{Shuryak:2021fsu},
 that its central potential can be derived from instanton-based model of the vacuum fields, with the benefit
 of also predicting spin-dependent nonperturbative forces,  at least comparable in magnitude to the perturbative ones.
 In \cite{Shuryak:2022thi} we  calculated  the instanton-induced quark-quark forces in various baryons,  and compared them
 to the empirical and lattice data.
 
 In our next publication we will continue along this line, focusing on interquark forces in baryons. While we start
 with generic three-Wilson-lines, the outcome are binary interactions only. We will also
 generalize  our approach to spin-dependent forces. All of that can also be further generalized to
 the recently discovered multi-quark hadrons.

\section{Light quark baryons in the CM frame}
The physics of baryon resonances is a very broad field. It is also rather old. It  started in  the 1950's with discoveries of
 the first $N^*$ and $\Delta$ resonances. The QCD-based quark models are also about 50 years old.
The systematic analysis of the data are in the Particle Data Tables,
for recent review see e.g. \cite{Gross:2022hyw}. Basically, it appears that the ``missing resonance"
problem is solved, and the states belonging to the ground, second and third shells are all
 observed, with quantum numbers for nearly all of them determined.

So, the reader may think that all of the main questions in this field are answered,
and doubt that
anything useful (rather than new experimental data) can be added to it. In fact, the situation is far
from that. The structure and especially the physical origin of the quark-quark forces
are not yet understood, especially of the spin-dependent part. Lattice studies  of  the static spin-dependent potentials are sketchy,  and in so far reduced to either $\bar q q$ channel or diquarks. These issues are now discussed again,
in view of the recent renaissance in the hadronic spectroscopy, due to the discovery of many new multi-quark hadrons. 

%In the first paper of our recent sequence \cite{I} we argued that (novel) model of the instanton ensemble
%leads to quite reasonable static potential, as well as spin-spin ones. In the paper \cite{IV} we 
%calculated the $qq$ forces in baryons, in the same approach. 

The question whether spin-dependent splittings of baryons are due to perturbative one-gluon
exchange,  or instanton-induced 't Hooft Lagrangian has been discussed in \cite{Shuryak:1988bf}.
Another proposed mechanism, 
via pion exchanges between quarks \cite{Glozman:1997fs}, appears as a higher order effect of the t Hooft Lagrangian,
as the pions themselves are mostly bound by it.  (We will not include those, trying to avoid a double counting.) In this vain, it is worth mentioning also, the solitonic construction based on the large number of colors limit~\cite{Zahed:1986qz}, which has been also extended to the exotics via the holographic principle~\cite{Liu:2021tpq,Liu:2021ixf}.

In this work we would like to start with another re-analysis of the situation, with a
phenomenological analysis of the data, mostly of positive parity (ground shell) and negative
parity (second shell)  light baryons  and discuss whether the existing data can accommodate certain
matrix elements of all spin-depending operators, namely spin-spin, spin-orbit, tensor as well as 
instanton-induced 't Hooft operator.

 In order to see the importance of various spin-dependent forces in action, it is $not$ sufficient to look at 
 the lowest shell baryons, $N$ and $\Delta$, as there is basically just one mass difference between them.
  To reach a quantitative understanding of the magnitude of the spin-dependent forces,
 we need to look at more states. An important step  in this direction was made by Isgur and Karl (IK) \cite{Isgur:1978wd} who focused on
 the P-shell negative parity  baryons. The Jacobi coordinates $\vec\rho,\vec\lambda$
 are antisymmetric and symmetric under permutation of the $1-2$ quarks, which leads
 naturally to the use of 
 blocks with the same $\rho,\lambda$ structure also for the orbital, spin and isospin parts of the 
 wavefunctions~\cite{Close:1979bt}. 
 This observation was used extensively in the work of IK and their followers.
 IK focused on hyperons with a strange quark, for which the wave function with 12-symmetry was sufficient.
The light quark baryons were then constructed as certain limits from the combination of hyperons, with the strange quark mass
reduced to  the light  ones. 
%not-fully-symmetrized wave functions. 
%Like for mesons, for which $L=1, p-shell$ states had provided a window into spin-orbit and tensor force
%actions, the same role for baryons  is played by negative parity states.  Let me remind the reader the setting, focusing on the 
%excited nucleons. 

Unlike Isgur-Karl, we decided  not to deal with extra parameters related with strangeness, and  focus entirely on
light   quark states. As it is well known,
negative parity $u,d$ baryons belonging to the second  $L=1$ (or P-shell) have seven  states,
 five $N^*$ and two $\Delta^*$. Five masses of the $N^*$ and two matrix elements of 
 mixing ($J=3/2$ and $J=1/2$ pairs with different spins $S$) are
 seven  inputs.
 We will use the inverse logic  adopted by most spectroscopists: instead of formulating a model and
 then comparing its prediction with various data, we decided to 
  start with  estimates of the {\em phenomenological values}  of the matrix elements of several
 contributing operators.  With the structure of the wavefunctions at hand, and spin-orbit-isospin structure of the operators known, one can write masses as linear combinations of matrix elements. There is enough information to fix
 those uniquely.

 \subsection{Excited nucleons, negative parity} 
We recall that that the spin states of  three quarks have $2^3$ states: four of those belong to the $S=3/2$ case, and the remaining two pairs of states 
have spin $S=1/2$ but different symmetries under permutations.  Adding orbital momentum $L=1$ to the former set
leads to  $J=5/2,3/2,1/2$ $N^*$ states. In the latter case two $S=1/2$ structures need to be combined
with two isospin $I=1/2$ structures, to get  a single combination with the correct permutation symmetry: this leads to 
another pair of $J=3/2,1/2$ of $N^*$.   Two pairs of $N^*$ from those two families, with the same $J$, are intermixed
by the tensor forces.

We start our phenomenological discussion with Table \ref{tab_masses}, where the current experimental masses for the P-shell states are recorded. We also compare these empirical masses with the original predictions by Isgur and Karl~\cite{Isgur:1978wd}. 

\begin{table}[b]
\caption{The second shell baryons made of light quarks. The original Isgur-Karl predictions~\cite{Isgur:1978wd}  are compared to the experimental masses from RPP. The last column shows the masses after ``unmixing" (see text).   }
\begin{center}
\begin{tabular}{|c|c|c|c|}
 \hline
 states $J^P$ & Isgur-Karl & experiment  & unmixed  \\ 
  \hline
 $ N^*_{1/2^-}$ & 1490 &1535 & 1567.3, $S=1/2$\\
  $ N^*_{1/2^-}$ & 1655 & 1650 & 1617.7 $S=3/2$\\
  $ N^*_{3/2^-}$ & 1535 & 1520 &  1521.97 $S=1/2$ \\
   $ N^*_{3/2^-}$ & 1745 & 1700&  1698.0 $S=3/2$\\
    $ N^*_{5/2^-}$ & 1670 & 1675 &  $S=3/2$ \\
%    $ \Delta^*_{1/2^-}$ & 1600 & & $S=1/2$ \\
%              $ \Delta^*_{3/2^-}$ &  1665 &  & $S=1/2$ \\
        \hline
\end{tabular}
\end{center}
\label{tab_masses}
\end{table}%

There are also mixing angles of the $S=3/2$ and $1/2$ states determined from the decays: those are listed
in the RPP reviews as being $\theta_{S1/2}=-32^o,\theta_{S3/2}=6^o$.
The relation of the energies of the mixed states to the unmixed ones is
\begin{equation} 
M_{\pm}= \frac 12 \bigg( M_1+M_2\pm \sqrt{4 H_{mix}^2+(M_1-M_2)^2} \bigg)
\end{equation}
and the mixing angle to the non-diagonal mixing matrix element is
\be 
tan\,\theta={2 H_{mix} \over M_1-M_2-\sqrt{4 H_{mix}^2+(M_1-M_2)^2}}
\ee

We may use  the values of the observed masses $M_{\pm}$, and the observed mixing angles to  derive the ``unmixed masses" $M_{1,2}$ with fixed total spin $S$, which we give in the last column
in the Table \ref{tab_masses}. Similarly, the numerical values of two mixing matrix elements  are
\ba \label{eqn_mixing_values}
H_{mix}(J=1/2)&=&51.7\, MeV \nonumber \\ H_{mix}(J=3/2)&=&-18.7\, MeV
\ea
Below we will use five masses to get five matrix elements, and compare the value of the tensor 
matrix element to these two mixing matrix elements.

\subsection{Permutation symmetry $S_3$ and three quark wavefunction}

Our first task is to clarify the use of $S_3$ for  the construction of baryon orbital-spin-isospin wavefunctions,
which can be made uniquely. 
Afterwhich, we will use phenomenology to evaluate
the matrix elements of all pertinent operators.

%  We do so first, before we calculate such matrix elements from available models.
%Permutation symmetry fixes wave function uniquely, and, with modern technical tools, there is no difficulty to
% deal with those.

%Proton $8_M$ WF from the 56-plet}
For completeness,  we start with  well known  basic facts.
Fermi statistics require antisymmetry over all permutations of quarks. With antisymmetric color wave function $\sim \epsilon_{ijk}$, the remainder of the wavefunction needs to be totally symmetric
\bea
{\rm color}_A\,\times \big({\rm space}\,\times\,\,{\rm flavor}\,\times\,{\rm spin}\big)_S
\eea
%the remainder of the wave function .
For orbital momenta $L_i=0$ (S-shell) states, the traditional classification follows
the representations of the spin-flavor  $SU(6)$ group of the nonrelativistic quark model 
\bea
6\otimes (6\otimes 6)=&&6\otimes (21_S\oplus 15_A)\nonumber\\
=&&(56_S\oplus 70_M)\oplus (20_A\oplus 70_M)
\eea
The nucleon
 belongs to the mixed symmetry $8_M$ octet, which is part of the symmetric 56-plet.
 As we do not include strange hyperons, flavor $SU(3)$ is reduced to the isospin $SU(2)$ group. 
 This helps to make spin and isospin notations more similar.
 
  The three quarks coordinate vectors are compressed into two Jacobi coordinates \bea \label{eqn_Jacobi}
\vec\rho=\frac 1{\sqrt{2}} \vec r_{12}\qquad \lambda=\frac{1}{\sqrt{6}}(\vec r_{13}+\vec r_{23})
\eea
with $\vec r_{ij}=\vec r_i-\vec r_j$ and 3D vectors assumed.  As noted by Isgur and Karl~\cite{Isgur:1978xj},  simple  symmetry properties of  $\varphi_{LM}^{\rho, \lambda}$ under the permutation group, suggest a convenient
set of basis functions, both in spin and isospin, out of which the baryon wavefunctions can be constructed.
Most of the details are given in Appendix \ref{sec_permutations}, with only  the main ideas presented here.

 To compose the states with proper symmetry under the permutation group $S_3$, we recall that this group 
is composed of 6 elements
\bea
\label{PERM}
P_{i=1, ...,6}=I, (12), (13), (23), (123), (132)
\eea
It will be enough to enforce the symmetry under (12) and (23) permutations. Let us add that there are three Young tableau's
for this group, symmetric, antisymmetric and mixed (two boxes in the top line and one below).

 The Jacobi coordinates (\ref{eqn_Jacobi})
% \bea \rho&=&(\vec r_1-\vec r_2)/\sqrt{2} \nonumber \\
% \lambda&=&(\vec r_1+\vec r_2-2\vec r_3)/\sqrt{6}
% \eea
% traditionally used, e.g.  in \cite{Isgur:1978xj}, 
 transform under permutations as follows
\bea
\label{eqn_perm_12}
{[P_2=(12)]}
\begin{pmatrix}
\rho\\
\lambda
\end{pmatrix}
&=&
\begin{pmatrix}
-1&0\\
0 & 1
\end{pmatrix}
\begin{pmatrix}
\rho\\
\lambda
\end{pmatrix}\nonumber\\
{[P_4=(23)]}
\begin{pmatrix}
\rho\\
\lambda
\end{pmatrix}
&=&
\begin{pmatrix}
\frac 12&\frac{\sqrt 3}2\\
\frac{\sqrt{3}}2& -\frac 12
\end{pmatrix}
\begin{pmatrix}
\rho\\
\lambda
\end{pmatrix}
\ea  
Thus $\rho,\lambda$ doublet transformation under (12) is simply antisymmetric and symmetric, but it is of 
mixed symmetry under the (23) permutations. Therefore, the construction of the states out of the $\rho$ or $\lambda$-type blocks (as done e.g. by
 Isgur and Karl~\cite{Isgur:1978wd}) allows for the easy removal of the 
 contributions that 
are not (12) symmetric.
 However, it is not enough: one needs to enforce full $S_3$ symmetry.
The corresponding matrices of (23) permutation for two and three quark is required (see Appendix),  ultimately
fixing the wavefunctions uniquely.

Let us start with the (well known) proton state.   In the litterature one can find it in several  forms, using either products of blocks with different symmetries, or a
sets of ``monoms" (basic states):
\begin{widetext}
\bea  \label{eqn_N_monoms}
&& \bigg| J= \frac 12,J_z=\frac 12, I_z= \frac 12 S_z=I_z= \frac 12 \bigg\rangle_{p_M^+} \nonumber \\
&=&{1 \over \sqrt{18}}\big[(\uparrow \downarrow \uparrow-\downarrow \uparrow \uparrow) (udu-duu) + (\uparrow \uparrow \downarrow-\uparrow \downarrow\uparrow)(uud-udu) + ( \uparrow \uparrow \downarrow-\downarrow \uparrow \uparrow)(uud-duu)\big] \nonumber  \\
&=& {1 \over \sqrt{18}}\big[ 2 (u^\uparrow d^\downarrow u^\uparrow)+2(d^\downarrow  u^\uparrow  u^\uparrow) +2(u^\uparrow  u^\uparrow d^\downarrow)-(d^\uparrow u^\downarrow u^\uparrow)-(u^\downarrow d^\uparrow u^\uparrow) \nonumber \\ &-&
(u^\uparrow d^\uparrow u^\downarrow) -(d^\uparrow u^\uparrow u^\downarrow)- (u^\downarrow u^\uparrow d^\uparrow)
-( u^\uparrow u^\downarrow d^\uparrow)
\big] 
 \eea
\end{widetext}
Its spin and isospin components  have mixed symmetry, but taken together, they are  symmetric under all permutations.
(Among its classic applications are the neutron-to-proton ratio of magnetic moments $-2/3$, etc.)
But in order to derive the wavefunctions for the baryon excitations, we need a more systematic approach.

There are $2^3=8$ spin (or isospin) states of three quarks. Four of those correspond to total spin $S=3/2$,
and four other to $S=1/2$. It is convenient to split them according to their symmetry under (12).
Examples of such  doublets are  the spin$-\frac 12$ states
\bea 
S_{\frac 12\frac 12}^\rho=&&\frac 1{\sqrt{2}}(\uparrow\downarrow-\downarrow\uparrow)\uparrow\nonumber\\
S_{\frac 12\frac 12}^\lambda=&&-\frac 1{\sqrt{6}}(\uparrow\downarrow\uparrow+\downarrow\uparrow\uparrow-2\uparrow\uparrow\downarrow)
\eea  \label{eqn_S12}
Similar isospin-$\frac 12$ states are defined as
\bea
F_{\frac 12\frac 12}^\rho=&&\frac 1{\sqrt{2}}(ud-du)u\nonumber\\
F_{\frac 12\frac 12}^\lambda=&&-\frac 1{\sqrt{6}}(udu+duu-2uud)
\eea
The $S=\frac 32$ (or $I=\frac 32$) are fully symmetric 
%\begin{widetext}
\bea \label{eqn_S_sym}
S_{\frac 32 m}^S= \bigg(\uparrow\uparrow\uparrow, \frac 1{\sqrt 3}(\uparrow\uparrow \downarrow + {\rm perm.}), \nonumber \\
\frac 1{\sqrt 3}(\uparrow\downarrow \downarrow + {\rm perm.}), \downarrow\downarrow\downarrow\bigg)
\eea
%\end{widetext}

To streamline the states  with $S_3$ symmetry $R=S,A,M$ (symmetric, anti-symmetric, mixed), we introduce the spectroscopic  notation
$$\big| LSJm\big\rangle_{X_R^P}$$
with the hadronic label $X^P_R$  for $X=p, \Delta, ...$ with parity $P=\pm$, and $L,S,J,J_z=m$ for orbital, spin and total angular momentum
with projection $J_z=m$, respectively. 

The simplest baryon $\Delta$ have  $S=I=\frac 32$ and both spin and isospin wave functions are symmetric $S^S$
constructions (\ref{eqn_S_sym})
\bea
\bigg|0\frac 32\frac 32m\bigg\rangle_{\Delta_S^+}=C_A\varphi_{00}S^S_{\frac 32 m}F^S_{\frac 32}
\eea
The proton ground state WF should have spin-isospin $S=I=\frac 12$, so it should be constructed out of (\ref{eqn_S12}).
There are four combinations.
Two of them $S^\rho F^\lambda,  S^\lambda F^\rho$ are asymmetric under (12) and should be rejected.
Two others are symmetric: any combinations of those is symmetric under (12). To fix the wave function uniquely,
we need to calculate their transformation under (23) permutations. As shown in the Appendix, the only combination
symmetric under both (12) and (23) is
\bea \label{eqn_proton_wf}
\bigg|0 \frac 12\frac 12 m\bigg\rangle_{p_M^+}\sim {\varphi_{00}}\frac 1{\sqrt{2}}(S_{ \frac 12 m}^\rho F_{\frac 12}^\rho+S_{\frac 12 m}^\lambda F_{\frac 12}^\lambda)
\eea
where the radial wave function $(\rho^2+\lambda^2)$ depends on 6d ``hyperdistance" $\sim (\rho^2+\lambda^2)$
%as well as the $\Delta-$isobar state spin-isospin $S=I=\frac 32$

Let us now proceed to the more general case of excited baryons, with the orbital part of the wave function
included.
In general, there could be angular functions depending on the angles of both vectors $\vec \rho,\vec \lambda$, but
for the P-shell we are interested in,  it is either $L_\rho=1,L_\lambda=0$ or $L_\lambda=1,L_\rho=0$.
So, for the
 spatial WFs $\varphi_{LM}^{\rho, \lambda}$ with orbital content of mixed symmetry  under $S_3$, we define
%The simplest excited P-state of the proton carries one unit of orbital angular momentum in the $\rho,\lambda$-coordinates
\bea
\label{ZMX}
\varphi^\rho_{1m}=&&(\rho_-, \sqrt{2}\rho_z, -\rho_+)\varphi_{00}\equiv z_m^\rho\varphi_{00}\nonumber\\
\varphi^\lambda_{1m}=&&(\lambda_-, \sqrt{2}\lambda_z, -\lambda_+)\varphi_{00}\equiv z_m^\lambda\varphi_{00}
\eea
with $\rho_\pm=(\rho_1\pm i\rho_2)$  and  $\lambda_\pm=(\lambda_1\pm i\lambda_2)$. Up to module
of the two vectors, they are the standard angular functions $Y_1^m$ with $m=-1,0,+1$.

They combine with the possible spins to give 5 nucleons negative parity proton excited states,
\bea
J^P=L^P\oplus  S^P&=&1^-\otimes \bigg({\frac 12}^+,  {\frac 32}^+\bigg)\nonumber\\
&\rightarrow& 2\times \bigg( {\frac 12}^-, {\frac 32}^-\bigg), {\frac 52}^-
\eea
i.e. two ${\frac 12}^-$, two ${\frac 32}^-$ and one ${\frac 52}^-$.
Yet to construct the wavefunctions, we need to include also the isospin in a proper way.
Modulo color, the baryon wavefunction has three components, orbital, spin and isospin. (Not to be confused
with the number of quarks: those will be there for any quark number.) As shown in Appendix A,
three objects are combined together into a unique combination (\ref{eqn_sym_in_3}).\\

\subsubsection{3-quarks with  $S=\frac 32$ with $J=\frac 52, \frac 32, \frac 12$}
The totally antisymmetrized wavefunction for 3 quarks with the maximally streched spin are
%$\bigg|L, S, J, J_z \bigg\rangle$
\bea
%p_m\bigg[{\frac 12}^-,{\frac 32}^-,{\frac 52}^-\bigg]=
\bigg|1 \frac 32\frac 52 \frac 52\bigg\rangle_{p_M^-}=
C_A \,S_{\frac 32\frac 32 }^S\,\frac 1{\sqrt{2}}(F_{\frac 12}^\rho \varphi_{11}^\rho+F_{\frac 12}^\lambda \varphi_{11}^\lambda)\nonumber\\
\eea
\begin{widetext}
\bea
\bigg|1 \frac 32\frac 32 \frac 32\bigg\rangle_{p_M^-}&=&C_A 
\bigg(\sqrt{\frac 35}\,S_{\frac 32\frac 32 }^S\,\frac 1{\sqrt{2}}(F_{\frac 12}^\rho \varphi_{10}^\rho
+F_{\frac 12}^\lambda \varphi_{10}^\lambda)-\sqrt{\frac 25}\,S_{\frac 32\frac 12 }^S\,\frac 1{\sqrt{2}}(F_{\frac 12}^\rho \varphi_{11}^\rho
+F_{\frac 12}^\lambda \varphi_{11}^\lambda)\bigg)\nonumber\\
\bigg|1 \frac 32\frac 12 \frac 12\bigg\rangle_{p_M^-}&=&C_A 
\bigg(-\frac 1{\sqrt 2}\,S_{\frac 32\frac 32 }^S\,\frac 1{\sqrt{2}}(F_{\frac 12}^\rho \varphi_{1-1}^\rho
+F_{\frac 12}^\lambda \varphi_{1-1}^\lambda)+\frac 1{\sqrt 3}\,S_{\frac 32\frac 12 }^S\,\frac 1{\sqrt{2}}(F_{\frac 12}^\rho \varphi_{10}^\rho
+F_{\frac 12}^\lambda \varphi_{10}^\lambda)\nonumber\\
&&\qquad -\frac 1{\sqrt{6}}\,S_{\frac 32 -\frac 12 }^S\,\frac 1{\sqrt{2}}(F_{\frac 12}^\rho \varphi_{11}^\rho
+F_{\frac 12}^\lambda \varphi_{11}^\lambda)\bigg)
\eea
with $S_{\frac 32\frac 32}^S=\uparrow\uparrow\uparrow$ the totally symmetric 3-quark spin $\frac 32$ state. 

In general, the addition of spin $S$ states with orbital functions to a particular $J$,    requires a sum of
kinematically possible states, $m=J_z=m_S+m_L$, using the standard Clebsch-Gordon coefficients
\bea
\label{SPIN32}
\bigg|1 \frac 32 J m\bigg\rangle_{p_M^-}= \sum_{m_S}
{\bf C}^{J m}_{1m_L\frac 32 m_S}\,\bigg[
C_A \,S_{\frac 32 m_s}^S\,\frac 1{\sqrt{2}}(F_{\frac 12}^\rho \varphi_{1m_L}^\rho+F_{\frac 12}^\lambda \varphi_{1m_L}^\lambda)\bigg]
\eea
Note that we use convention for the Clebsch-Gordon coefficients in terms of Wigner 3-j symbol
\bea
{\bf C}^{Jm}_{Lm_LSm_s}=(-1)^{S-L-m}\sqrt{2J+1}
\begin{pmatrix}
L & S & J\\
m_L & m_s & -m
\end{pmatrix}
\eea
 
%\end{widetext}

\subsubsection{3-quarks in spin $\frac 12$ with $J=\frac 32, \frac 12$}
%Similarly, we have
%\\
%\\
%\begin{widetext}
%{\bf 3-quarks in spin $\frac 12$:}
The maximally streched J-states for fixed $L, S$ are
\bea
\label{SPIN12}
%p_m\bigg[{\frac 12}^-,{\frac 32}^-\bigg]=
\bigg|1 \frac 12\frac 32 \frac 32\bigg\rangle_{p_M^-}&=&C_A \,
\frac 1{\sqrt{2}}\bigg(F_{\frac 12}^\rho \frac 1{\sqrt{2}}(\varphi_{11}^\rho S_{\frac 12\frac 12}^\lambda +\varphi_{11}^\lambda S_{\frac 12\frac 12}^\rho)+
F_{\frac 12}^\lambda \frac 1{\sqrt{2}}(\varphi_{11}^\rho S_{\frac 12\frac 12}^\rho-\varphi_{11}^\lambda S_{\frac 12\frac 12}^\lambda)\bigg)\nonumber\\
\bigg|1 \frac 12\frac 12 \frac 12\bigg\rangle_{p_M^-}&=&C_A \bigg(
\sqrt{\frac 23}\frac 1{\sqrt{2}}\bigg(F_{\frac 12}^\rho \frac 1{\sqrt{2}}(\varphi_{11}^\rho S_{\frac 12-\frac 12}^\lambda +\varphi_{11}^\lambda S_{\frac 12-\frac 12}^\rho)+
F_{\frac 12}^\lambda \frac 1{\sqrt{2}}(\varphi_{11}^\rho S_{\frac 12-\frac 12}^\rho-\varphi_{11}^\lambda S_{\frac 12-\frac 12}^\lambda)\bigg)\nonumber\\
&&\qquad
-\sqrt{\frac 13}\frac 1{\sqrt{2}}\bigg(F_{\frac 12}^\rho \frac 1{\sqrt{2}}(\varphi_{10}^\rho S_{\frac 12-\frac 12}^\lambda +\varphi_{10}^\lambda S_{\frac 12-\frac 12}^\rho)+
F_{\frac 12}^\lambda \frac 1{\sqrt{2}}(\varphi_{10}^\rho S_{\frac 12-\frac 12}^\rho-\varphi_{10}^\lambda S_{\frac 12-\frac 12}^\lambda)\bigg)\
\bigg)
\eea
\bea
%p_m\bigg[{\frac 12}^-,{\frac 32}^-\bigg]=
\bigg|1 \frac 12\frac 32 \frac 32\bigg\rangle_{\Delta_M^-}=C_A \,
 F^S_{\frac 32}\,\frac 1{\sqrt{2}}\bigg(S_{\frac 12\frac 12}^\rho\varphi_{11}^\rho+S_{\frac 12\frac 12}^\lambda\varphi_{11}^\lambda\bigg)
\eea
The lower J-states given $L,S$ follow  by Clebsch-Gordoning,
%For the odd parity $J=\frac 12, \frac 32$ shells in the proton, 
\bea
\label{MX1}
%p_m\bigg[{\frac 12}^-,{\frac 32}^-\bigg]=
\bigg|1 \frac 12 J m\bigg\rangle_{p_M^-}=\sum_{m_S}{\bf C}^{J m}_{1m_L\frac 12 m_s}\,\bigg[C_A \,
\frac 1{\sqrt{2}}\bigg(F_{\frac 12}^\rho \frac 1{\sqrt{2}}(\varphi_{1m_L}^\rho S_{\frac 12  m_s}^\lambda +\varphi_{1m_L}^\lambda S_{\frac 12 m_s}^\rho)+
F_{\frac 12}^\lambda \frac 1{\sqrt{2}}(\varphi_{1m_L}^\rho S_{\frac 12 m_S}^\rho-\varphi_{1m_L}^\lambda S_{\frac 12 m_s}^\lambda)\bigg)\bigg]\nonumber\\
\eea
and for the odd parity $J=\frac 12, \frac 32$ shells in the isobar,
\bea
%p_m\bigg[{\frac 12}^-,{\frac 32}^-\bigg]=
\bigg|1 \frac 12J m\bigg\rangle_{\Delta_M^-}=\sum_{m_S}{\bf C}^{Jm}_{1m_L\frac 12 m_s}\,\bigg[C_A \,
 F^S_{\frac 32}\,\frac 1{\sqrt{2}}\bigg(S_{\frac 12 m_s}^\rho\varphi_{1m_L}^\rho+S_{\frac 12 m_s}^\lambda\varphi_{1m_L}^\lambda\bigg)\bigg]
\eea
\end{widetext}

The method for deriving the orbital-spin-isospin wavfunctions out of
the $\rho,\lambda$ blocks of mixed permutation symmetry,  is basically known. Their 
use using Jacobi-like combination was carried by Isgur and Karl.
However, the
explicit wavefunctions in the current literature, are not in a  standard form useful for applications to multiquark states. Since while working on this
paper we have developed them, we will  explain the simple rules on how it was done in Mathematica, and present a full set of explicit wavefunctions
in Appendix B.%we explicitly show that the mixed symmetry terms in (\ref{MX1}) satisfy the correct permutation symmetry.

\section{$J^P=2^+$ excited states} \label{sec_L2}
Representation theory of angular momentum  tell us that one
 can construct positive parity excited nucleon states with 
 $J^P$ assignments
% Using (\ref{PDX}) we can construct positive parity excited nucleon states with
% $J^\pi$ assignments
 $$ L^\pi\otimes S=2^+\otimes \frac 12, \frac 32= \bigg({\frac 32}^+, {\frac 52}^+\bigg),  \bigg({\frac 12}^+, {\frac 32}^+,  {\frac 52}^+, {\frac 72}^+\bigg) $$
% $$J^\pi=L^\pi\otimes S=1^+\otimes \frac 12= {\frac 12}^+, {\frac 32}^+$$
However, the explicit construction of the wave functions symmetric under quark permutations needs further attention. These wave functions were discussed e.g. in~\cite{Capstick:2000qj}. However the full treatment is better achieved 
using the representations of the $S_3$ permutation group, to be discussed below in 
Appendix \ref{sec_permutations}. 

Since orbital part of the 
wavefunction for $J^P=2^+$ are symmetric tensors constructed
out of coordinate vectors $\rho^i,\lambda^i$, one has three options for this
$$\rho^i\rho^j,\lambda^i\lambda^j,
(\rho^i\lambda^j+\lambda^i\rho^j)$$ The first two are symmetric under [12], the last is antisymmetric. Under [23] permutation their transformation is involved under
(\ref{eqn_XX}). To get totally symmetric wavefunctions, 
they need to be supplemented by spin and isospin wavefunctions with appropriate 
symmetries.

For the total spin $S=\frac 32$, the corresponding spin wavefunctions are symmetric (e.g. $\uparrow \uparrow \uparrow$),  and we need to apply the $S_3$ representation with three
objects (two coordinates and isospin) $X^{A1}X^{A2}X^{A3}$ with binary indices $A=\rho,\lambda$. We have shown that it is a single combination (\ref{eqn_sym_in_3}). 

For the total spin $S=\frac 12$,  there are two options $S^\rho,S^\lambda$. 
 The most straightforward way to construct the wavefunctions is via
building symmetric representation of $four$ spinor-like objects $$X^{A1}X^{A2}X^{A3}X^{A4}$$ with  binary indices $A=\rho,\lambda$. 
Their total number is $2^4=16$, half of them symmetric and half antisymmetric
under [12] permutations. The $16\times 16$ matrix (not shown) of [23] permutations
have 8 symmetric and 8 antisymmetric eigenvectors. In the Appendix,  we show that there are $two$ 
 symmetric under $all$ permutations,
 which can be rewritten as
\be \lambda^i \lambda^j S_\lambda I_\lambda +\rho^i \lambda^j S_\lambda I_\rho +\lambda^i \rho^j S_\rho I_\lambda +\rho^i \rho^j S_\rho I_\rho 
\ee
\be \lambda^i \lambda^j S_\rho I_\rho -\rho^i \lambda^j S_\lambda I_\rho -\lambda^i \rho^j S_\rho I_\lambda +\rho^i \rho^j S_\lambda I_\lambda  
\ee
Their linear combinations generate tensor excited states with spin $S=\frac 12$.
For those with fixed $J,J_z$ one needs, as usual, to calculate the Clebsch-Gordan coefficients.

In the Appendix we have pushed the method one step further, to 5 $\rho,\lambda$ blocks. Out of 32 states, we found 3 symmetric eigenstates. 
\section{Splittings of negative parity nucleons }
%	\fbox{Edward based on summary_1.nb Aug.9}\\
	The details of the operators and calculation methods are  presented below
 and in the Appendices, but we start by summarising the results. 
 
 All the matrix elements contain 6-dimensional integrals over   $\vec \rho,\vec \lambda$ coordinates which are split into two sets, 4d angular and two remaining radial
\be \int\,\bigg({d\Omega_\rho \over 4\pi}{d\Omega_\lambda \over 4\pi}\bigg)\,(\rho^2 d\rho \lambda^2 d \lambda ) ... \ee
times pertinent matrix elements.
 The wavefunctions consist of a common 6d spherically symmetric function $\phi_{00}(R), R^2\equiv \rho^2+\lambda^2$,
 times various orbital parts. For the P-wave baryons under consideration, the orbital parts are linear in $\vec\rho,\vec\lambda$. The matrix elements receive multi-contributions from the different parts of the symmetrized wavefunctions, as detailed in the Appendix. Four angular integrals will be performed
 in matrix elements  to follow, while integrals over the moduli $\rho,\lambda$
 remains undone, as no assumptions about $\phi_{00}(R)$ is made.
 We introduce a shorthand notations for such integrals 
  \begin{equation} \langle \hat V(\rho,\lambda) \rangle \equiv \int  \int d\rho d\lambda \cdot  \rho^2  \lambda^2 |\phi_{00}|^2 V(\rho,\lambda) 
  	\end{equation}
For example, in such notations the normalization integral ($\hat V=\hat 1$)  ``averages"
 the orbital wave function squared, namely
  	 \be  \label{eqn_averaging} 
    \langle (\rho^2+\lambda^2) \rangle= \int  \int d\rho d\lambda \cdot  \rho^2  \lambda^2 |\phi_{00}|^2 (\rho^2+\lambda^2)\ee
This combination will appear in denominators of all terms.  

The masses of the $N^*_{J}$ excited nucleons have some overall 
constant (not written below)  plus contributions of four spin operators, e.g. spin-spin, spin-orbit, tensor and 't Hooft terms. The resulting mixing matrices in the P-sub-shells are

%the calculations are in summary_1.nb
	
\begin{widetext}
\ba \label{eqn_results}
{\mathbb M}_{\frac 52}'=
{3\over 4}{\langle  (\rho^2+\lambda^2)  V_S(\sqrt{2}\rho)\rangle\over \langle (\rho^2+\lambda^2) \rangle}
		+3{\langle \rho^2 V_{LS}(\sqrt{2}\rho) \rangle \over \langle (\rho^2+\lambda^2) \rangle } 
	-{1\over 5} {\langle \rho^4 V_T(\sqrt{2}\rho) \rangle \over \langle (\rho^2+\lambda^2) \rangle}
 %+6(1-a) G_{tH} {I_{\lambda} \over \langle (\rho^2+\lambda^2) \rangle } 
\ea

\bea
{\mathbb M}_{\frac 32}' &=&
\begin{pmatrix}
	3/4 & 0\\
	0 & -3/4
\end{pmatrix} \,{\langle  (\rho^2+\lambda^2)  V_S(\sqrt{2}\rho)\rangle\over \langle (\rho^2+\lambda^2) \rangle} +
\begin{pmatrix}
	-2 & -\sqrt{5/2}\\
	 -\sqrt{5/2} & 1
\end{pmatrix} \,{\langle  \rho^2  V_{SL}(\sqrt{2}\rho)\rangle\over \langle (\rho^2+\lambda^2) \rangle} 
\nonumber \\ &+&
\begin{pmatrix}
	{4 \over 5} & {1\over \sqrt{10}}\\
{1\over \sqrt{10}} &	0
\end{pmatrix} \, { \langle  \rho^4  V_T(\sqrt{2}\rho) \rangle \over \langle (\rho^2+\lambda^2) \rangle } 
+
\begin{pmatrix}
0 & 0\\
0 & (1+3a) 
 \end{pmatrix} { G_{tH}  \langle\lambda^2\rangle \over 8 \sqrt{2} \pi \langle (\rho^2+\lambda^2) \rangle }
\\
{\mathbb M}_{\frac 12}'&=&
\begin{pmatrix}
	3/4 & 0\\
	0 & -3/4
\end{pmatrix} \,{\langle  (\rho^2+\lambda^2)  V_S(\sqrt{2}\rho)\rangle\over \langle (\rho^2+\lambda^2) \rangle}  -
\begin{pmatrix}
	5 & 1 \\
1 	& - 2
\end{pmatrix} \,{\langle  \rho^2  V_{SL}(\sqrt{2}\rho)\rangle\over \langle (\rho^2+\lambda^2) \rangle } 
\nonumber \\ &+&
\, \begin{pmatrix}
	- 1 &  -1\\
	-1 & 0
\end{pmatrix} \,  {\langle  \rho^4  V_T(\sqrt{2}\rho)\rangle\over \langle (\rho^2+\lambda^2) \rangle }  
+ \begin{pmatrix}
0 & 0\\
0 & 1+3a 
 \end{pmatrix} { G_{tH}  \langle\lambda^2\rangle \over 8 \sqrt{2} \pi \langle (\rho^2+\lambda^2) \rangle }
\eea
The rows and columns of the $2 \times 2$ matrices, correspond to states with $S=\frac 32, \frac 12$, respectively, 
\bea
\label{DEG1}
\mathbb M_J=
\begin{pmatrix}
\langle S=\frac 32|\mathbb V_{S+T+SL+V_{TH}}|S=\frac 32\rangle  & \langle S=\frac 32|\mathbb V_{S+T+SL+V_{TH}}|S=\frac 12\rangle \\
\langle S=\frac 12|\mathbb V_{S+T+SL+V_{TH}}|S=\frac 32\rangle  & \langle S=\frac 12|\mathbb V_{S+T+SL+V_{TH}}|S=\frac12\rangle 
\end{pmatrix}_J
\eea
Some of these matrices have nondiagonal (or mixing)
matrix elements, which means that the total spin $S$ is not a good quantum number of 
the observed states,

\end{widetext}
We now explain how the matrix elements are calculated. 
All operators used should be symmetric under quark permutations, as
are all the orbital-spin-isospin wavefunctions. In most cases we calculated the pair 12-interactions, and then multiplied the result by 3 (accounting 
for three pairs, 12, 13, 23). In Jacobi coordinates, the 12-distance  is $d_{12}=\sqrt{2}\rho$, so most potentials are fixed at that distance. 

The operators are defined as follows
\ba \label{eqn_operators}
H_{SS}^{12}&=& V_S(d_{12}) (\vec S_1 \vec S_2) \\
H_{LS}^{12}&=& V_{LS}(d_{12}){(S_1+S_2)^i\over 2} \rho^j \big({\partial \over i \partial \rho^k}\big) \epsilon^{ijk} \nonumber \\
H_{T}^{12}&=&V_{T}(d_{12})[(\vec S_1 \vec d_{12})(  \vec S_2 d_{12})-(\vec S_1 \vec S_2) d_{12}^2 ] \nonumber \ea
Since , the
    corresponding terms are proportional to $\langle \rho^2 V_{LS}(\sqrt{2}\rho)\rangle $. the numerical factors in front
  each contribution, stem from the convolution over all indices of operators, wavefunctions and 4d angular integrals.

The exception is the instanton-induced interaction taken in the local approximation, 
\be H_{tHooft}^{12}= G_{tH} \delta^3(\vec d_{12})(1-\vec \tau_1\vec \tau_2) (1-a \vec \sigma_1 \vec \sigma_2)
\ee
where $\vec \tau_i,\vec \sigma_i$ are Pauli matrices for  isospin and spin, respectively. The 't Hooft interaction, is quasi-local in nature, and for the 12-quark pair is  proportional to $\sim G_{tH}\delta^3(\vec d_{12})$ . As a result, the integral over $\vec\rho$ drops out, and only the integral over $\lambda$ remains,
so we define for it a lambda-only  averaging
 \begin{equation} I_{\lambda}\equiv  \int d\lambda \lambda^4   |\phi_{00}(\lambda)|^2  
 \end{equation}

\subsection{Extracting matrix elements from $N^*$ data}

The knowledge of the coefficients of the relevant operators for the five (unmixed) masses of  negative parity $N^*$, allows to fix the pertinent matrix elements. This
 immediately reveals one striking feature, already noted by Isgur and Karl~\cite{Isgur:1979be}: the  role of spin-orbit is an order of magnitude (or more) suppressed relative to the spin-spin and tensor force. The same observation
 follows for the 't Hooft operator which we tried to include: it is not improving the
 fit and its expectation value is within the error bars.  (For P-shell baryons, not so for the ground state nucleon).

%Putting $a=4/5$ (for three colors) one has 5 parameters, the spin-independent mass and 4 matrix elements.
%Using 5 available masses, we get the following values for matrix elements (in %$MeV$)
%\ba \langle H_0\rangle&=&1626.29, \\
%\langle V_{SS}\rangle&=& 64.92\,\,\, \langle V_{SL}\rangle=-3.35 \nonumber \\
%\langle V_{tensor}\rangle&=& 149.95\,\,\, \langle V_{'t Hooft}\rangle %=-3.51\nonumber
%\ea
%Since we get 5 parameters from 5 masses, these formulae reproduce all of them exactly.
%One can see that there are two large effects -- the spin-spin and the tensor -- and two small effects, both expected.
%Surprisinly small spin-orbit contribution has been  now we can put a number on this
%qualitative observation. Smallness of 
%'t Hooft interaction has also been expected in excited baryons, on the ground that instanton sizes are small,
%$\sim 0.3\, fm$,  and it is quasi-local force, while
%the p-shell baryonic wave function have zero values at the origin due to centrifugal potential, and relatively 
%large r.m.s. size. 

 In view of this we resorted to what we call "an optimized IK model",
which ignores LS and 't Hooft terms and keep only the spin-spin and the tensor forces. In this case, the fitted  matrix elements can be considered reliable.
\ba \label{eqn_optimized}
\langle H_0 \rangle= 1607. \, MeV \nonumber \\
%cSS -> 83.1745, cT -> 43.7254
{\langle \rho^2 V_{SS}\rangle \over \langle \rho^2+\lambda^2 \rangle }= 83.2\, MeV, \nonumber \\
{\langle \rho^4 V_{tensor}\rangle \over \langle \rho^2+\lambda^2 \rangle } =43.7 \, MeV
\ea
where $H_0$ is  the spin-independent part of the Hamiltonian, common to all five resonances.

\begin{figure}[h]
\begin{center}
\includegraphics[width=8cm]{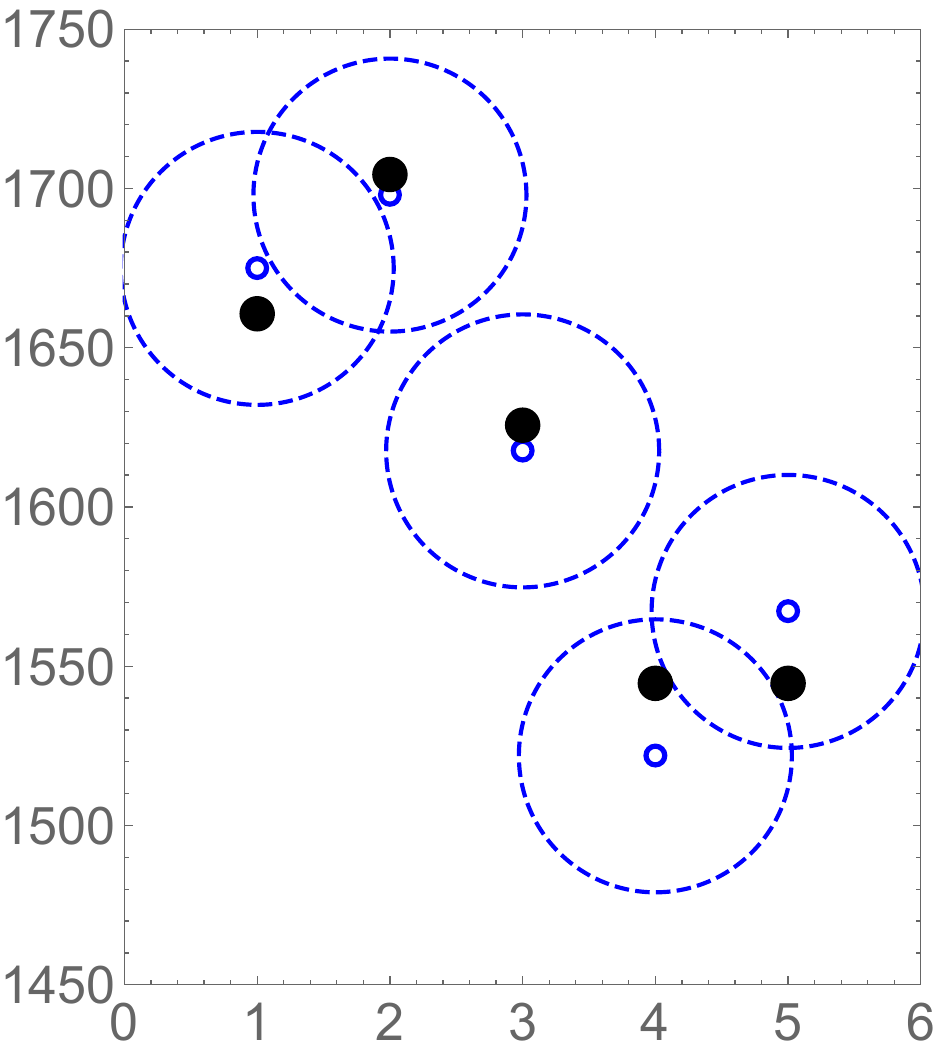}
\caption{The black dots are masses of negative parity P-shell baryons, from the "optimized IK model" with  fitted radial matrix elements  (\ref{eqn_optimized}), in the order $|J=\frac 52>,|J=\frac 32,S=\frac 32>,|J=\frac 12,S=\frac 32>,|J=\frac 32,S=\frac 12>,|J=\frac 12,S=\frac 12>.$ The small blue circles show the experimental pole positions. They are surrounded by large dashed blue circles, with radius of the order of  typical half width $\frac 12\Gamma$. }
\label{fig_5nucs}
\end{center}
\end{figure}

Our only assumptions so far are using spin-dependent forces to the first order,
and ignoring near-threshold effects. 
(Isgur and Karl made additional assumptions, such as a Gaussian S-wavefunction,
used to evaluate them. )
The quality of the overall description of masses can be seen  in Fig.\ref{fig_5nucs}. One can see that the agreement of this model with the data is 
very good. In  all cases, it is  significantly better  than the half-widths of these resonances
(which provides a scale for the ignored threshold effects).

 As  ``external tests" of the accuracy of such a model, 
we use the empirical mixing matrix elements (\ref{eqn_mixing_values}). 
The observed ratio for the  magnitude of the mixing matrix elements in the tensor case is 
\be {H_{mix}(J=1/2)
\over H_{mix}(J=3/2)}\approx -2.76 \ee
If only the tensor operator is left, the ratio of its coefficients $-\sqrt{10}=-3.16$  is sufficiently close to it. 
%Yet declaring complete success of the 
%(first order) theory would be premature: 
The value from the fitted tensor matrix element to the splittings in Fig.\ref{fig_5nucs}, we have the value (\ref{eqn_optimized}) or $43.7\, MeV$. 
 This is to be compared to $51.6\, MeV$ from the mixing matrix element in the $J=\frac 12$-shell pair 
 as discussed above.
 Hence, some other effects, not yet accounted for here, contribute at the 
level of $\sim 8 \, MeV$. A similar conclusion can also be inferred from
the deviations between the observed and fitted masses seen in Fig.\ref{fig_5nucs}.
These deviations should be due to higher order corrections not yet
accounted for.  They are reasonably small compared to the splitting themselves $\sim 100\, MeV$ as well as to
 $\Gamma/2\sim 50\, MeV$ used as the maximal possible shifts due to threshold locations.  

%%%%%%%%%%%%%%%%%%%%%%%%%%%%

%%\subsubsection{Mixing angles in the P-shell}
%The diagonalization of (\ref{DEG1}) yields the eigen-equation
%\begin{widetext}
%\bea
%\label{DEG1X}
%\begin{pmatrix}
%\langle \frac 32|\mathbb V_{S+T+SL}|\frac 32\rangle  & \langle \frac 32|\mathbb V_{S+T+SL}|\frac 12\rangle \\
%\langle \frac 12|\mathbb V_{S+T+SL}|\frac 32\rangle  & \langle \frac 12|\mathbb V_{S+T+SL}|\frac12\rangle 
%\end{pmatrix}_J
%{\begin{pmatrix}
%\alpha_{\frac 32}\\
%\alpha_{\frac 12}
%\end{pmatrix}}^{\pm}_J
%=\Delta^J_\pm 
%{\begin{pmatrix}
%\alpha_{\frac 32}\\
%\alpha_{\frac 12}
%\end{pmatrix}}^{\pm}_J
%\eea
%\end{widetext}
%with the physical mixed P-states 
%\bea
%N^-_{\frac 32 J}=\alpha^+_{\frac 32 J}N_{\frac 32 J}+\alpha^+_{\frac 12 J}N_{\frac 32 J}\nonumber\\
%N^-_{\frac 12 J}=\alpha^-_{\frac 32 J}N_{\frac 32 J}+\alpha^-_{\frac 12 J}N_{\frac 32 J}
%\eea
%with the energy splitting $\Delta^J_+-\Delta^J_-$ and mixing angles
%$${\rm tan}(\theta^\pm_J)=\frac{\alpha^\pm_{\frac 12 J}}{\alpha^\pm_{\frac 32 J}}$$
%
%
%

\section{Spin-dependent interactions in baryons}
Having summarized so to say  the``applied aspects" of the calculations,
let us return to the theory. Before discussing the spin-dependent effects, few words about the
 spin-$independent$ potentials. Those are expressed in terms of Wilson-line correlators, or complicated nonlocal correlations of (Euclidean) vacuum gauge fields. They can be computed numerically on the lattice,
or evaluated in some models of gauge fields in the vacuum. An instanton-based model we used for quarkonia in  \cite{Shuryak:2021fsu}, and claimed that it 
can explain most of the confining potential, at least till the relevant distances $< 1\, fm$. For baryons, there are correlators of
three Wilson lines, see their calculation in the instanton model \cite{Shuryak:2022thi} and detailed comparison with
lattice results. In a recent paper \cite{Miesch:2023hjt} we extended such calculations for the three
Wilson lines (for baryons) and four ones (for tetraquarks). In  the latters, the  hyperditance
potential was shown to describe the distances between the three $cc \bar c\bar c$ states
recently discovered, and interpreted as $1S,2S,3S$ states. Application to $uuu=Delta^{++}$ S-shell states were also successful. No attempt so far
were made along these lines to describe baryons or multiquark states made
of different flavor quarks. In such cases we need to go beyond  the hyperdistance
approximation.

The generic spin-dependent five potentials were defined 
in~\cite{Eichten:1980mw}, in terms of  Wilson-lines dressed by pertinent  gauge field strengths.  Specific relations between them and the central potential, 
follow for one-gluon exchange. For example, the spin-spin
potential is famously a Laplacian of the central potential, a delta function
for the Coulomb force.  Similar  but distinct relations follow for  the instanton-induced potentials, mostly constrained by self-duality of the instanton fields. 
However, if a significant fraction of the vacuum fields are due to the overlapping instantons and antiinstantons, there could be significant corrections. 
 The spin-dependent interactions in the baryons and tetraquarks due to instantons can be evaluated, and we hope to report on them in our subsequent publications.

In this section we only include some general comments  related to the definition of the 
operators. The specifics of the calculation for fixed states, are summarized in the Appendix.  

%%%%%%%%%%%%%%%%%%%%%%%%%%%%%%%%%%%%%%%%%%%%%%%%%%%%%%%%%%%%%%%%%%%% 
\subsection{Spin-spin and tensor interactions}
We use the same strategy in the evaluation of all the spin interactions. More specifically, the 12-interactions are folded over 
the pertinent wavefunctions, 
and the overall results are 
multiplied by 3 by symmetry.
The results for
spin-spin forces agree with the much simpler ``quantum-mechanical" method.

For the ``unmixed" states  with fixed total spin of three quarks, $S=\frac 32$ or $S=\frac 12$, 
we have $\vec S=\vec S_1+\vec S_2+\vec S_3$, so that
\be \langle \vec S_1\vec S_2+\vec S_1\vec S_3+\vec S_3\vec S_2\rangle =\frac 12 \big(S(S+1)-9/4\big)
\ee
hence $\pm \frac 34$ for the cases of interest. 
The l.h.s. is symmetric under quark permutations, and is the only form for the spin-spin forces. 
Of course, each term is convoluted with potentials, that depend on the relative distance, e.g. $$(\vec S_1 \cdot \vec S_2)V_{SS}(d_{12}), \,\,\,\vec d_{12}=\sqrt{2}\vec \rho$$
the average of which multiplied by 3. 

The tensor operator we used is given in (\ref{eqn_operators}) in a standard way, and also evaluated it using the 12-pair. Note that in the tensor,  we are using  the
actual distance vector $\vec d_{12}$ rather than its unit vector version. Hence the 
extra $2\rho^2$ in the matrix element (to be accounted for in the potential $V_T$).

\subsection{Spin-orbit operator}
 Two-body systems (mesons) have one orbital momentum $L$ and one total spin $S$, so the natural spin-orbit force is proportional to $(\vec S \vec L)$. For three quarks
 there are $\vec L_\rho$ and $\vec L_\lambda$ orbital and three different quark spin structures. In general, it is still not all, since there can be 
 products of one coordinate vector times momentum of another one.  
 The requirement of permutation symmetry helps.  Suppose  the spin-dependent potentials are binary (two-body). With this in mind, the spin-orbit contribution is a sum of binaries 
\begin{widetext}
\bea
\mathbb V_{SL}(1,2,3)&=&\sum_{i<j=1,2,3}^3
%\frac 1{m_Q^2}\frac 1{r_{ij}}\frac{dV_C(r_{ij})}{dr_{ij}}
V_{SL}(d_{ij})
\big[(r_{ij}\times p_i)\cdot \sigma_i-(r_{ij}\times p_j)\cdot \sigma_i)\big]\nonumber\\
&\equiv& \mathbb V_{SL}(1,2)+V_{SL}(1,3)+V_{SL}(2,3)
\eea
\end{widetext}
Recall that the spin-orbit relation to the central potential is suppressed $V_{SL}\sim {1 \over m_Q^2}$.
This is the case of all relativistic corrections.

For the instanton-induced 
spin-orbit potential we have $$V_{SL}= \frac 1{m_Q^2}\frac 1{r_{ij}}\frac{dV_C(r_{ij})}{dr_{ij}} $$in terms of the  instanton-induced central  potential. To recast it in terms of the Jacobi coordinates, we recall that
\bea
r_{12} =\sqrt 2 \rho\qquad r_{13,23} = \frac 1{\sqrt 2}(\rho^2+3\lambda^2\pm \sqrt 3\rho\cdot\lambda)^{\frac 12}\nonumber\\
\eea
and the corresponding momenta 
\bea
\begin{pmatrix}
p_1\\
p_2\\
p_3
\end{pmatrix}
=
\begin{pmatrix}
\frac 1{\sqrt 2}&\frac 1{\sqrt 6} & 0\\
-\frac 1{\sqrt 2}& \frac 1{\sqrt 6} & 0\\
0& -\frac 2{\sqrt 6} & 0
\end{pmatrix}
=
\begin{pmatrix}
p_\rho\\
p_\lambda\\
0
\end{pmatrix}
\eea
In particular, for the 12-spin-orbit contributi we have
%\begin{widetext}
\bea
\label{SL1}
\mathbb H_{SL}(1,2)&=&V_{SL}(d_{12}) 
\bigg(\bigg({\sigma_1+\sigma_2 \over 2}\bigg)\cdot (\rho\times p_\rho) \nonumber \\
&+& \frac 1{\sqrt 3}(\sigma_1-\sigma_2)\cdot (\rho\times p_\lambda)\bigg)
\eea
Here the first contribution is the standard spin-orbit contribution, proportional to $L_\rho$, rotation of a 12-pair. 
The second term is also symmetric under  12-permutation, but has a vector product of the coordinate $\rho$ and momentum $p_\lambda$,
so it is neither two-body nor even an angular momentum. Fortunately, we found
that all the second term contributions vanish after angular integration, basically because $\vec \rho$ and $\vec \lambda$ have independent directions. Therefore we do not need to include this complicated part of the operator.

\subsection{ 't Hooft induced interactions}
The instanton-induced forces follow from the fermionic zero modes discovered by 't Hooft. Those are 
antisymmetric in flavors, and form generically 6-fermion operators of the $\bar u u \bar d d \bar s s$ flavor structure.
If we ignore strangeness (technically, by substituting $\bar s s$ part by its vacuum average value), the interaction reduces to a 4-fermion
$ud$-interaction. It can be shown to be of 
 the form
% \begin{widetext}
 \bea
 \label{XTHX}
 &&\mathbb V_{TH}(1,2,3)=\nonumber\\
 &&\sum_{i<j=1,2,3}\, V_{TH}(r_{ij})\,(1-\tau_i\cdot\tau_j)(1-a\,\sigma_i\cdot\sigma_j)\nonumber\\
 \eea
%\end{widetext}
% V_{Hooft}(r_{ij}) (1-\vec \tau_i\vec \tau_j) (1-a \vec \sigma_i\vec \sigma_j) \ee
in the ultra-local approximation, with 
$$V_{TH}(r_{ij})\rightarrow -\frac 14|\kappa_2| \,A_{2N}=-\frac 14|\kappa_2| \,\frac{2N_c-1}{2N_c(N_c^2-1)}$$
and
\bea
a=4B_{2N}=\frac 1{(2N_c-1)}
\eea
The coefficient $a$ comes from terms with color matrices, and assuming that we discuss baryons and carry the average 
 over all instanton color orientations, we get
  $a=\frac 15$ for $N_c=3$ and $a\rightarrow 0$ for  $N_c\rightarrow \infty$.
 
 The first factor in (\ref{XTHX}) with isospin matrices $\vec \tau$, is a projector on the flavor singlet states (zero in $\Delta$). So it contributes to the $N-\Delta
$ splitting in the S-shell. There is also no problem to evaluate this operator  over the  five $N^*$ states 
in the P-shell. In the quasi-local approximation $V_{tHooft}\sim \delta^3(d_{12})$,
 of the $\sim \rho^2,\lambda^2$ contributions arising from the orbital wavefunctions, only the $\lambda^2$ remains.
 %we only kept
% the latter terms. 

% \subsection{Spin induced interactions}

%The spin-spin and tensor interactions can be addressed similarly
%\bea
%\mathbb V_{S+T}(1,2,3)=\sum_{i<j=1,2,3}^3\bigg(\frac 1{12 %m_Q^2}\nabla^2V_C(r_{ij})\sigma_i\cdot\sigma_j+
%\frac 1{m_Q^2}\bigg(\frac 1{r_{ij}}V_C^\prime(r_{ij})-V_C^{\prime\prime}%(r_{ij})\bigg)
%(\sigma_i\cdot \hat r_{ij}\sigma_j\cdot \hat r_{ij}-\frac 13 \sigma_i\cdot %\sigma_j)\bigg)\nonumber\\
%\eea
%Because of permutation symmetry of the nucleon states, the matrix elements $X=TH, %S,SL, T$, satisfy
%\bea
%\langle N_A|\mathbb V_{X}(1,2)|N_B\rangle =\langle N_A|P_{23}\mathbb V_{X}(1,2)P_{23}|N_B\rangle =\langle N_A|\mathbb V_{X}(1,3)|N_B\rangle\nonumber\\
%\langle N_A|\mathbb V_{X}(1,3)|N_B\rangle =\langle N_A|P_{12}\mathbb V_{X}%(1,3)P_{12}|N_B\rangle =\langle N_A|\mathbb V_{X}(2,3)|N_B\rangle
%\eea

%\end{widetext}
%In the center of mass frame, the matrix elements of the spin-induced interactions in states with $\rho\lambda$-coordination can be obtained by reduction using the Wigner-Eckart theorem. This is more subtle in the light front frame where SO(3,1) symmetry is reduced to SO(2,1) symmetry.
%

%%%%%%%%%%%%%%%%%%%%%%%%%%%%%%%%%%%%%%%%%%%

\section{Baryon wave functions on the light front} 
There are many differences between the  CM and the LF  wavefunctions induced simply by kinematics. 
For example, the former uses nonrelativistic (or semi-relativistic~\cite{Capstick:1985xss}) descriptions, not well justified 
for light quarks. The latter (in the form developed in our previous papers \cite{Shuryak:2023siq,Shuryak:2022wtk,Shuryak:2022thi,Shuryak:2021hng,Shuryak:2021fsu,Shuryak:2023fjj}) does not need this assumption,
it takes the same form, from light $u,d,s$ to heavy $c,b$ quarks. Furthermore, it naturally provides an oscillator-like behavior for transverse
momenta, which translate into a $linear$ dependence of the baryon {\em mass squared},  on the number of excitation quanta.

However, on the LF manifest rotational symmetry is lost, as transverse and longitudinal motions
are treated differently. The states are no longer classified by their total angular momentum $J$, orbital $L$ or total spin $S$. 
Only the helicity projection of those, $J_z,L_z$ and $S_z$, can be used.  As we will see 
in this case, the importance of permutation symmetry (as it was illustrated in previous sections and Appendix A) is even broader.

\begin{widetext}
A simple Fock state representation of the spin-up proton wavefunctions on the LF was originally given in~\cite{Ji:2003yj}
\bea \label{eqn_Ji_WF}
|p^\uparrow\rangle_{L_z=0}&=&\int d[1,2,3]\, (\psi_1(1,2,3)+i\epsilon^{\alpha\beta}k_{1\alpha\perp}k_{2\beta\perp} \psi_2(1,2,3))\nonumber\\
&&\times\frac{\epsilon^{ABC}}{\sqrt{6}}\,b^{A\dagger}_{u\uparrow}(1)\bigg(b^{B\dagger}_{u\downarrow}(2)b^{C\dagger}_{d\uparrow}(3)
-b^{B\dagger}_{d\downarrow}(2)b^{C\dagger}_{u\uparrow}(3)\bigg)|0\rangle\nonumber\\
|p^\uparrow\rangle_{L_z=+1}&=& \int d[1,2,3]\, (k^+_{1\perp}\psi_3(1,2,3)+k^+_{2\perp}\psi_4(1,2,3))\nonumber\\
&& \times\frac{\epsilon^{ABC}}{\sqrt{6}}
\bigg(b^{A\dagger}_{u\uparrow}(1) b^{B\dagger}_{u\downarrow}(2)b^{C\dagger}_{d\downarrow}(3)
-b_{d\uparrow}^{A^\dagger} (1)b^{B\dagger}_{u\downarrow}(2) b^{C\dagger}_{u\downarrow}(3)\bigg)|0\rangle\nonumber\\
|p^\uparrow\rangle_{L_z=-1}&=& \int d[1,2,3]\, {k}^-_{2\perp}\psi_5(1,2,3)\,
\frac{\epsilon^{ABC}}{\sqrt{6}}\,b^{A\dagger}_{u\uparrow}(1)\bigg(b^{B\dagger}_{u\uparrow}(2)b^{C\dagger}_{d\uparrow}(3)
-b^{B\dagger}_{d\uparrow}(2)b^{C\dagger}_{u\uparrow}(3)\bigg)|0\rangle\nonumber\\
|p^\uparrow\rangle_{L_z=+2}&=& \int d[1,2,3]\, k^+_{1\perp}k^+_{3\perp}\psi_6(1,2,3)\,
\frac{\epsilon^{ABC}}{\sqrt{6}}\,b^{A\dagger}_{u\downarrow}(1)\bigg(b^{B\dagger}_{d\downarrow}(2)b^{C\dagger}_{u\downarrow}(3)
-b^{B\dagger}_{u\downarrow}(2)b^{C\dagger}_{d\downarrow}(3)\bigg)|0\rangle
\eea
where generically $b^{C\dagger}_{fs}(1)=b^{C\dagger}_{fs}(x_i, k_{i\perp })$, $k_\perp^\pm =k_x\pm k_y$,  and
\bea
d[1,2,3]=(2\pi)^3\delta^3\bigg(\sum_{i=1}^3k_{i\perp}\bigg)\delta\bigg(1-\sum_{i=1}^3x_i\bigg)
\prod_{i=1}^3\frac{dx_i\,dk_{i\perp}}{\sqrt{2x_i}(2\pi)^3} \nonumber\\
\eea
\end{widetext}
The fermionic creation operators anti-commute, and once multiplied by the color indices $\epsilon^{ABC}$, they generate 12 terms. 

The first term $\psi_1$ 
in the upper raw (times the second raw) above  reproduces the
correct permutation-symmetric S-wave function of the proton, in the form we already used in the upper raw
of (\ref{eqn_N_monoms}). 
 The expressions for the proton  with $L_z=\pm1$ are linear in (transverse) momentum, and  obviously $\psi_1$
 refer to a spin-up proton with {\em negative parity}. The 3 $\psi_{3,4,5}$ LF wavefunctions  refer to part of the  nucleon P-shell, 
we discussed above. They do  not cover all the 5 P-nucleon states we derived, as they have
 different wave functions, with different permutation content.  Also, their longitudinal momentum content under permutation symmetry 
 is not specified.
%One more issue here is: why
 %only some (not all) momenta are present, no longitudinal ones and without clear permutation symmetry.
 %We will address those issues below.
  Finally, the second term in the first raw and the last one are quadratic in momenta. So, the 2 $\psi_{2,6}$  wavefunctions correspond
 to D-shell excitation or their admixtures, with positive parity. Their permutation structure is
 incomplete. This issue has been addressed in the CM above in section    \ref{sec_L2}, and will be extended below to the LF.

% Indeed, as it is well known and was reminded above, symmetric spin-isospin wave function of the nucleon (\ref{}) is made
% uniquely from mix symmetry parts, as requiered by the $S_3$ symmetry. As we reminded above, for $L>0$ states with nontrivial orbital 
% wave functions, constructing wave functions with proper Fermi symmetry gets much more involved,
 %but still can be done  uniquely with proper construction of higher $S_3$ representations. 
%  p-shell negative parity nucleon excitations 
% with $J=3/2$ and $J=1/2$  have two states each, distinguished by their permutation symmetries, and mixed only
% by spin-dependent forces.
% 

%\subsection{p-shell baryons on the light front}
To clarify the argument, 
 let us first recall again the situation in the CM frame, returning to our main  example of   the P-shell  nucleons. 
 Physical states in the CM formulation have spherical symmetry, and therefore
  fixed total angular momentum $J$. They are specific combinations of all possible values of the orbital and spin helicities $L_z,S_z$. For example,
% For example, let us following construction by \cite{Ji} for $J_z=1/2, P=-1$ baryons.
 % It is clear that those must be superpositions of the following three helicity states are available
 \bea 
 &&| N^* J_z=1/2,P=-1> =\nonumber\\
 &&+f_1 | L_z=1, S_z=-1/2>\nonumber \\ 
&& + f_0 | L_z=0, S_z=1/2> \nonumber\\ && +f_{-1} | L_z=-1, S_z=3/2>  \nonumber
 \eea
 are superpositions of states with all three $L_z$ values, with some coefficients $f_{L_z}$.
  For states with fixed $J$  those are defined  by standard Clebsch-Gordon rules,
 which for this particular example prescribe the
  coefficients $f_1,f_0,f_{-1}$  to  be 
 \ba \label{eqn_with_Clebsches}
 \sqrt{6} &&\bigg( {Y_{11}(\theta,\phi) \over 2\sqrt{15}}, {Y_{10}(\theta,\phi)\over \sqrt{10}}, {Y_{1-1} (\theta,\phi)\over 2\sqrt{5}} \bigg), \nonumber \\ 
  2&&\bigg( {Y_{11} (\theta,\phi) \over \sqrt{10}},{Y_{10} (\theta,\phi)\over 2\sqrt{15}},-{2 Y_{1-1}(\theta,\phi)\over \sqrt{15}}\bigg),\nonumber \\ 
  \sqrt{2} &&\bigg(  {Y_{11}(\theta,\phi)  \over 2},-{Y_{10}(\theta,\phi)\over 
  \sqrt{6}} ,{Y_{1-1}(\theta,\phi) \over   2 \sqrt{3}} \bigg)
 \ea
 for $J=5/2,3/2,1/2$, respectively, and $J_z=1/2$. Note that these three vectors are indeed mutually orthogonal and normalized. 
  Under the zeroth order Hamiltonian,
 they all have the same energy, and only the spin-dependent forces create the observed splittings.These states are defined in the CM frame as eigenstates of the total angular momentum $\vec J$, but this option is not available in the LF formulation.

 Furthermore, as emphasized above, Clebsching
  is not  enough for baryons. Their  wave unctions are $linear$ in coordinates,  which bring in their negative parity. There are 6  $\vec \rho,\vec \lambda$, and one expects 6 basic orbital wave functions.
  The same number of basis states needs to be defined in the LF formulation. 
   The transverse polarization component $L_z=0$, proportional to  the $longitudinal$ coordinates/momenta 
 $\rho_z,\lambda_z$ in the CM frame, needs to be redefined. As we will show below, they should be substituted
 by solutions on the triangle, of $\rho$ and $\lambda$-type respectively 
\be Y_{10}(\theta_\rho)  \rightarrow D^\rho,\,\,\,\,                               Y_{10}(\theta_\lambda)  \rightarrow D^\lambda \ee

Also, the spin $ S_z=\pm \frac 12$  structure  in fact exists in three forms, 
the fully symmetric one corresponding to $S=\frac 32$, and two more mixed with $\rho$-like and $\lambda$-like
permutation properties. The same statement applies to isospin wave function.  Constructing the correct combinations is not a trivial task, and it was the subject of the preceding sections.

%  Proceeding to angular momentum basis, the only change
% is that transverse coordinates gets grouped into $ \rho_x\pm i\rho_y, \lambda_x\pm i\lambda_y$ 
% combinations with $\phi$ angles. $L_z=0$ components have longitudinal coordinates
% $\rho_z,\lambda_z$. There are still 6 functions, $L_z^\rho,L_z^\lambda=1,0,-1$.

\subsection{Light front wave functions  for negative parity baryons} \label{sec_LF_long}
Now we switch to the main subject of this work, the construction of the corresponding wave functions on the light front (LF),
with a focus on the P-shell baryons. 
Four out of the six coordinates corresponding to the transverse 12-plane, remains unchanged. Since we 
proceed on LF in momentum representation, the angular functions with $L_z=\pm 1$ are proportional to either $k^\rho_1\pm i k^\rho_2$ or $k^\lambda_1\pm i k^\lambda_2$ .
The requirements of   12-permutation symmetry helps to eliminate many impossible combinations,
and to reduce the number of functions necessary.
In particular,  the spin-isospin-transverse functions should be superpositions of  {\em permutation-symmetric} blocks, such as 
$$   \big( e^{\pm i\phi_\rho}I_\rho+e^{\pm i\phi_\lambda}I_\lambda\big) S_{sym} $$ for p-shell nucleons with spin $\frac 32$  and 
  $$ \big( e^{\pm i\phi_\rho}S_\rho+e^{\pm i\phi_\lambda}S_\lambda\big) I_{sym}$$
  for $\Delta$ baryons. Other combinations are formed as in the CM frame.

The question remains what are the other two other basis states, with longitudinal momenta. Using Bjorken-Feynman
longitudinal momenta $fractions$ $x_i \in [0,1],i=1,2,3$ as variables and enforcing the kinematical constraint $x_1+x_2+x_3=1$, 
we will again use Jacobi variables to characterize them. We will refer to them by $\rho,\lambda$ without any indices. (They  should not be
 confused with the lengths of the coordinate vectors $\vec \rho, \vec \lambda$ used in
 previous sections.)
 
 As discussed in our paper \cite{Shuryak:2022thi}, the physical domain
in this case is the equilateral triangle, between three points at which one of the $x_i$ reaches unity and two others vanish,
\bea
&&(\rho, \lambda) \in \nonumber\\
&&{\rm Triangle}\bigg( \{0, -\sqrt{2 \over 3}\}, 
  \{ {1 \over \sqrt{2}}, {1 \over \sqrt{6}}\}, \{-{1 \over \sqrt{2}}, 
    {1 \over \sqrt{6}}\} \bigg) \nonumber
\eea
The role of the permutation group $S_3$ is now clearly seen,. In these notations, it is a set of reflections and rotations of this triangle.
%What are quantum numbers of corresponding wave function in respect to its generators we are going to discuss now.
%

Quantization on this equilateral triangle has been done in  \cite{Shuryak:2022thi}, which was achieved both analytically and numerically. 
We recall that 
that the confining
potential, after an ``einbine trick", can be made quadratic in coordinates. In the momentum representation we use the sum of the squared coordinates, to turn it Laplacian in  momenta $\rho,\lambda$. The analytic solutions follow in the form   of six waves,  interfering at the triangle boundaries using Dirichlet  (zero) boundary condition, see (54) of  \cite{Shuryak:2022thi}.
The spectrum of the Laplacian is given by two integers $m_L,n_L$ ($L$ refers  to the longitudinal directions).
\be \label{eqn_Laplace_eigenvalues}
\epsilon_{m_L,n_L}={8 \pi^2 \over 3^2}\big(m_L^2+n_L^2 - m_Ln_L\big) \ee
The single-degenerate solutions correspond to $m_L=2n_L$, with the two lowest already shown and discussed in  \cite{Shuryak:2022thi}.

\begin{figure}[htbp]
\begin{center}
\includegraphics[width=6cm]{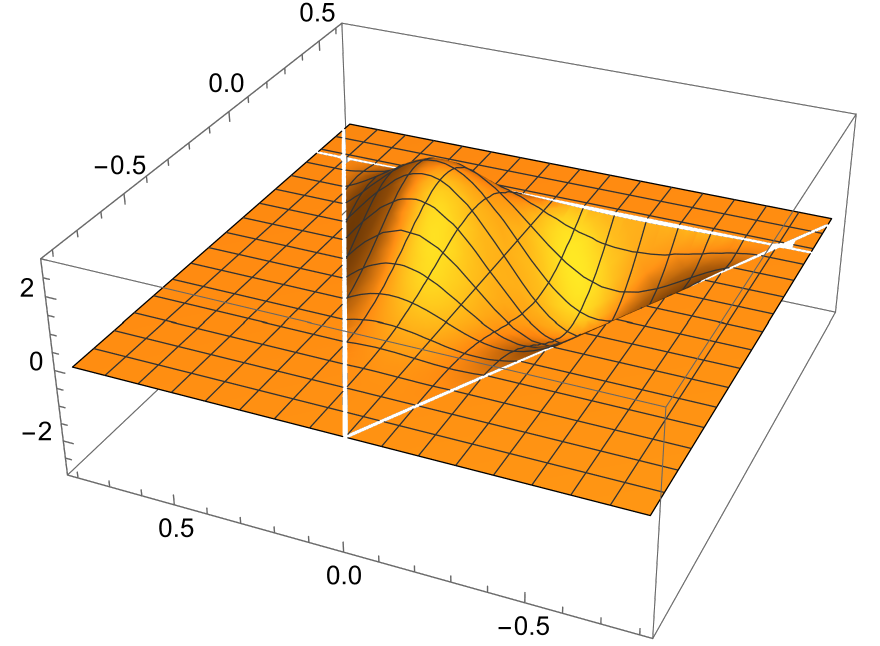}
\includegraphics[width=6cm]{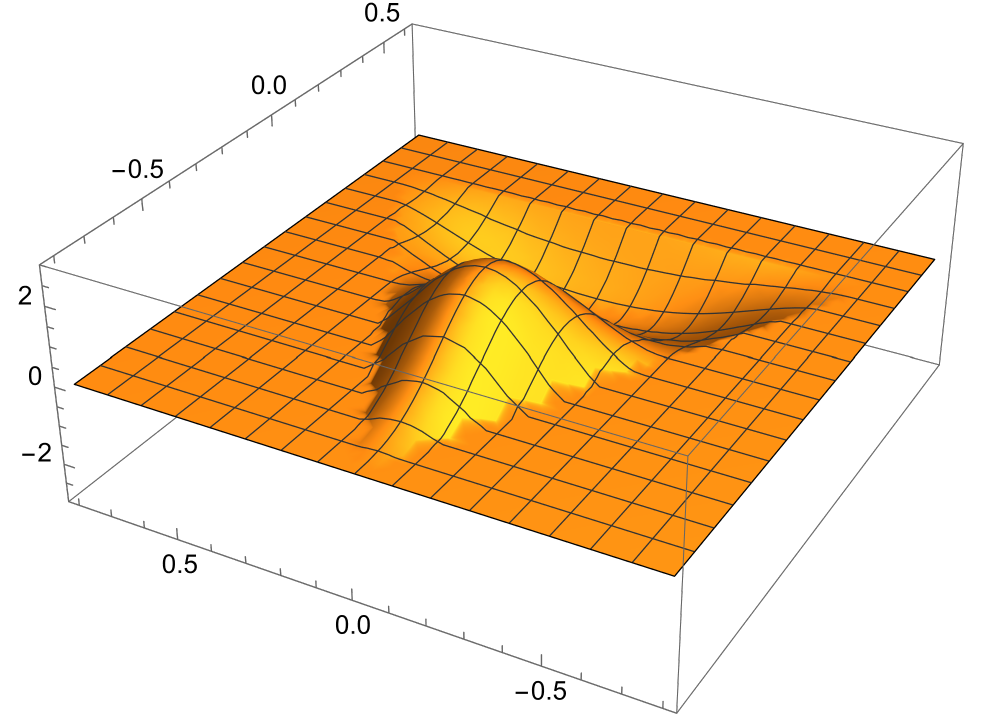}
\caption{The $D^\rho, D^\lambda$  solutions of the Laplacian on the equilateral triangle, for $m_L=3,n_L=1$ }
\label{fig_D_triangular}
\end{center}
\end{figure}

Now we  focus on the solutions with $m_L> 2n_L$, which are in fact $double$ degenerate 
\bea
\label{BER1}
&&^{D^\lambda}_{m,n}(\lambda, \rho)=\frac{4 }{L\,3^{\frac 34}}\bigg[{\rm cos}\bigg(\frac{2\pi(2m_L-n_L)\rho}{3L}\bigg)
{\rm sin}\bigg(\frac{2\pi n_L\tilde\lambda}{\sqrt{3}L}\bigg)\nonumber\\
&&-{\rm cos}\bigg(\frac{2\pi(2n_L-m_L)\rho}{3L}\bigg)
{\rm sin}\bigg(\frac{2\pi m_L\tilde\lambda}{\sqrt{3}L}\bigg)\nonumber\\
&&+{\rm cos}\bigg(\frac{2\pi(m_L+n_L)\rho}{3L}\bigg)
{\rm sin}\bigg(\frac{2\pi (m_L-n_L)\tilde\lambda}{\sqrt{3}L}\bigg)\bigg]\nonumber\\
&&{D^\rho}_{m,n}(\lambda, \rho)=\frac{4 }{L\,3^{\frac 34}}\bigg[{\rm sin}\bigg(\frac{2\pi(2m_L-n_L)\rho}{3L}\bigg)
{\rm sin}\bigg(\frac{2\pi n_L\tilde\lambda}{\sqrt{3}L}\bigg)\nonumber\\
&&-{\rm sin}\bigg(\frac{2\pi(2n_L-m_L)\rho}{3L}\bigg)
{\rm sin}\bigg(\frac{2\pi m_L\tilde\lambda}{\sqrt{3}L}\bigg)\nonumber\\
&&-{\rm sin}\bigg(\frac{2\pi(m_L+n_L)\rho}{3L}\bigg)
{\rm sin}\bigg(\frac{2\pi (m_L-n_L)\tilde\lambda}{\sqrt{3}L}\bigg)\bigg]\nonumber
\eea
with $\tilde\lambda=\lambda+L/\sqrt{3}$. Their   symmetry properties include (1-2) (or $\rho \rightarrow -\rho $ symmetry)
\bea
D^{\rho,\lambda}_{m,n} (\lambda, -\rho)=\pm D^{\rho,\lambda}_{m,n}(\lambda, \rho)
\nonumber
\eea
(Yes, the triangle has obvious triple symmetry by $120^o$ rotations, but those produce linear
combinations of these two solutions.)
 Those were called $D^c,D^s$ before, because the former includes  combinations only with
$ {\rm cos}(C_i\rho)$, and the latter  similar set of ${\rm sin}(C_i \rho)$. In the present paper, where permutation symmetry is central, we would like to rename them into $D^\lambda,D^\rho$, respectively.
Indeed, the former is even under 12-permutation, and the latter is odd for $\rho\rightarrow -\rho$, see Fig.\ref{fig_D_triangular}.
These are the solutions  of $\rho$ and $\lambda$ types, which on the LF are substitutes for zero orbital momentum components in the CM, $L_z^\lambda=0$ and $L_z^\rho=0$, essentially simple linear coordinates $\lambda_z,\rho_z$.

Since these are new solutions, let us discuss how the spin operators act on them. 
The spin-orbit interactions include orbital momenta such as $$L^i_\rho=\epsilon^{ijk} \rho^j ( i\partial/\partial \rho^k)$$
In the coordinate representation the momentum is a derivative, and in the momentum representation (we are using on the LF)  the coordinates are
 derivatives over momenta. 
Acting  by $L$ on the longitudinal function of longitudinal momentum $\rho\sim p^\rho_z$, the vector product can only get
other -- transverse-- components $p^\rho_{x,y}$ due to the first factor. Since those appear linearly, it means the
$L$ operation converts longitudinal functions into transverse ones. As a result, on the LF the angular momentum operator
mixes two transverse functions with one longitudinal, as it does in the CM description for the three angular functions in the P-shell.  

Another thing to notice is that the derivative in $L$ interchanges
${\rm cos}$ and ${\rm sin}$. With the appearance of $\rho$-
type transverse coordinates, we see that this operator does not change parity under  
12-permutation. Yet, the differentiation changes the function, so it is not just $D^\rho \leftrightarrow D^\lambda$ as 
there are additional admixtures of higher functions, with other $m_L,n_L$. 

Similar thing happens with the tensor force. In the CM formulation, the interaction generates nonzero non-diagonal matrix elements
for $L=0 \leftrightarrow L=2, L=1 \leftrightarrow L=3$ shells etc. The spin part in the tensor causes mixing of the $S=\frac 32$ and $S=\frac 12$ states, as  we discussed above.
Yet, since the coordinate tensor times the spin tensor is 
rotationally a scalar, $J$ remains unchanged. 
In the LF formulation, there is no rotational symmetry or $J$, and the tensor force mixes all $D^\rho, D^\lambda$ functions.

Let us now address the issue of parity. The standard $P$ mirror reflection, sign change of all coordinates,
makes odd-$L$ states with negative parity, which cannot be mixed with positive parity states.
On the LF the notion of parity is remedied by an
additional rotation by $180^o$, so that longitudinal momenta do not flip. In particular,
our equilateral triangle in the  $\rho,\lambda$ plane maps into itself. So, formally, the $D^\rho, D^\lambda$ functions
do not have parity as such.

 Yet these LF functions have certain parities under particle permutations. As one can see from Fig.\ref{fig_D_triangular},
 $D^\rho$ is odd under flipping of $\rho$ and $D^\lambda$ under flipping of $\tilde\lambda=\lambda+\sqrt{2/3}$
 (since they contain  the corresponding sin functions). Those are sufficient to define odd and even shells, which 
 cannot be mixed by any forces/operators preserving the permutation symmetry.

In sum, the six wave functions out of which the P-shell baryons are made on the LF,  consist of
four transverse ones, with ${\rm exp}(\pm i \phi_\rho), {\rm exp}(\pm i \phi_\lambda)$, complemented by two new $D^\rho, D^\lambda$ 
functions on the triangular domain for longitudinal momenta fraction. The P-shell baryons can be constructed
 out of these mixed-symmetry blocks by the same expressions as used above. 
 
 More explicitly, the  $S=\frac 32$ with  fixed $J_z$ the spin WF is symmetric, and the orbital and isospin ones are mixed symmetries
 of $\rho,\lambda$ types. We know that the full $S_3$ symmetric combination out of the two mixed blocks,  has
  the structure $X_1^\rho X_2^\rho +X_1^\lambda X_2^\lambda$. Therefore the {\it permutations-symmetric} wave functions are
  \begin{widetext}
  \bea
  \label{LFP1}
 | N^{*\uparrow}_{S=\frac 32,J_z} \rangle= \int d[1,2,3]\,C_A\,\sum_m 
  S^S_{\frac 32 (J_z-m)}\bigg(\bigg(A_m^{\frac 32 J_z}\,e^{im\phi_\rho}+B_m^{\frac 32 J_z}\,D_{3,1}^\rho\bigg)\,F_{\frac 12}^\rho
  +\rho\rightarrow\lambda\bigg)\bigg)
  \eea
  %\ba | N^{*\uparrow}_{S=3/2} \rangle&=&\int d[1,2,3]\sum_m  S_{S_z=J_z-m}^{sym}(s_1,s_2,s_3) \times \nonumber \\ 
 %  \big(A_m e^{i m \phi_\rho}  &+& B_m D^\rho_{3,1} )I^\rho(i_1,i_2,i_3)  + \nonumber \\ 
 %\big( A_m e^{i m \phi_\lambda}  &+& B_m D^\rho_{3,1} \big)I^\lambda (i_1,i_2,i_3)
%    \ea
and similarly for the $S=\frac 12$ and fixed $J_z$ we have
   \bea
   \label{LFP2}
&& | N^{*\uparrow}_{S=\frac 12,J_z} \rangle= \int d[1,2,3]\,C_A\,\sum_m \nonumber\\
 &&\times
 \bigg(F_{\frac 12}^\rho\bigg(\bigg(A_m^{\frac 12 J_z}\,e^{im\phi_\rho}+B_m^{\frac 12 J_z}\,D_{3,1}^\rho\bigg)S^\lambda_{\frac 12(J_z-m)}+\lambda\leftrightarrow \rho\bigg)
 +
 F_{\frac 12}^\lambda\bigg(\bigg(A_m^{\frac 12 J_z}\,e^{im\phi_\rho}+B_m^{\frac 12 J_z}\,D_{3,1}^\rho\bigg)S^\rho_{\frac 12(J_z-m)}-
 \rho\rightarrow\lambda\bigg)\bigg)\nonumber\\
  \eea
   \end{widetext}
We note that the two $L=\pm 1$ lines in (\ref{eqn_Ji_WF})
should be substituted by  the  5 states (\ref{LFP1}-\ref{LFP2}) for the P-shell $N^*$. The same applies to the 
2 $\Delta^*$.
The coefficients $A_m, B_m$ are not fixed kinematically in the absence of rotational symmetry. 
They are determined dynamically through the diagonalization of the LF Hamiltonian. Clearly, a
fine tuning of the parameters is likely needed to achieve the same mass spectrum for different $J_z$.
This is in sharp contrast with the CM where rotational symmetry fixes these coefficients as Clebsch-Gordon
coefficients as in (\ref{eqn_with_Clebsches}). 

%so in LF formulation $A_m,B_m$ one should find from diagonalizing the Hamiltonian. Certain parameter tuning would be needed if masses of different $J_z$ components be approximately the same. )

Finally, note that since the transverse part of the LF Hamiltonian  is an oscillator, we have $M^2_n\sim n_\perp$. The longitudinal 
 eigenvalues (\ref{eqn_Laplace_eigenvalues}) are quadratic in quantum numbers, yet after  minimization over the einbine parameter,
 they enter via $(m_L^2+n_L^2-m_L n_L)^{\frac 12}$. They are not linear, but relatively  close to linear. More specifically, the first three double-degenerate excitations, counted from
 from the ground state (n=2m=2), have energies 
 \be \sqrt{(m_L^2+n_L^2-m_L n_L)}-\sqrt{3}=0.913,1.87,2.85
 \ee
 to be compared to 1,2,3 for an oscillator. So, with a pertinent fit to the parameters,
 one should be able to get agreement between the masses for different $J_z$ components in the same $J$-multiplet, 
 with few percent accuracy. Similar accuracy is expected for  Regge phenomenology.  
 Furthermore,  if one would only compare states with the same $J_z$  and different $J$ (as we did for negative parity baryons in (\ref{eqn_with_Clebsches}))  such deviations are further reduced, as they only appear in the longitudinal component, a part of the wavefunction. So far,
 our discussion refers to  the bare Hamiltonian $H_0$. The   spin-dependent potentials 
 cause larger shifts , $\sim 10\%$ for the P-shell baryons, and we expect that this imperfection 
 in the longitudinal part of the spectrum on the LF,  will not seriously affect the spectroscopy.

 \subsection{Wavefunctions on LF for positive parity excitations}
 In the CM frame the D-shell baryons have wavefunctions $quadratic$ in coordinates, as we already discussed above. Each of the coordinates have 2 transverse and one longitudinal components, of $\rho$ or $\lambda$ types. So, in general, one may approach the problem asking for representations with $four$ 
 structures $X_1\otimes X_2 \otimes X_3\otimes X_4$, of $\rho$ or $\lambda$ types, and seek their
 totally symmetric combinations. It can be done along the same reasoning as for the  $X_1\otimes X_2 \otimes X_3$ case we addressed in appendix A. This case would correspond to
  the case of the $S=\frac 32$ symmetric in spin.

%\fbox{why do we need the next section? proton and negative parity nucleons are already constructed}
%\section{Proton WF with diquarks in 21-plet}
%To reorganize the ground state WF of the proton with manifest diquark content (Fierzing) we recall that in SU(6),
%the diquarks compose as
%\bea
%6\otimes 6=21_S\oplus 15_A
%\eea
%The diquarks in the symmetric $21_S$ correspond to the vector 6-plet plus scalar 3-plet combinations
%($3\times 6+1\times 3=21$).
%They can be used to compose the proton WF. In particular, the spin=isospin=$\frac 12$
%proton with spin-up, has a both scalar and vector diquark content of the form
%\begin{widetext}
%\bea
%p^\uparrow=C_A\bigg[\frac 1{\sqrt{2}}[ud]_Su^\uparrow +\frac 1{\sqrt{2}}\bigg(\sqrt{\frac 23}[uu]_Vd-\sqrt{\frac 16}[ud]_Vu\bigg)_f\bigg(\sqrt{\frac 23}1\downarrow -\sqrt{\frac 16}0\uparrow\bigg)_s\bigg]
%\eea
%or equivalently
%\bea
%p^\uparrow=C_A\bigg[\frac 1{\sqrt{18}}\bigg(3[ud]_Su^\uparrow+2[uu]^+_Vd^\downarrow-\sqrt{2}[uu]^0_Vd^\uparrow-\sqrt{2}[ud]^+_Vu^\downarrow+[ud]^0_Vu^\uparrow\bigg)\bigg]
%\eea
%\end{widetext}
%where $[fg]_{S,V}$ stand for scalar (S) and vector (V) diquark flavor combination.
%
%
%
%

\section{Leading twist proton DAs}
The simple classification of the lowest Fock states for a spin-up  proton state made of three constituent quarks, 
consists of the 6 LF wavefunctions in (\ref{eqn_Ji_WF}). 
As we noted earlier, this Fock state representation mixes
the ground and excited states of the nucleon with fixed 
spin $\frac 12$. Boosting the nucleon in its ground S-state,
should leave the nucleon in its ground S-state. Also the simple Fock state representation forgoes the issue of the center of mass and symmetrization,  which are important 
for the description of the nucleon S,P,D states as we have discussed.

This notwithstanding, the Fock state representation is useful for the characterization of generic spin-$\frac 12$ nucleon distribution amplitudes (DAs), in the leading twist approximation. More specifically, 
three quarks with spin-$\frac 12$ can combine in four  different ways  to  form a proton with spin up $p^\uparrow$, i.e.
\bea
p^\uparrow =p_0^\uparrow +p_{+1}^\uparrow +p_{-1}^\uparrow +p_{+2}^\uparrow 
\eea
with the spin-orbital arrangements
\bea
p^\uparrow_0&=&\bigg(\uparrow \downarrow\uparrow=+\frac 12\bigg)+(L_z=0)\nonumber\\
p^\uparrow_{+1}&=&\bigg(\downarrow \downarrow\uparrow=-\frac 12\bigg)+(L_z=+1)\nonumber\\
p^\uparrow_{-1}&=&\bigg(\uparrow \uparrow\uparrow=+\frac 32\bigg)+(L_z=-1)\nonumber\\
p^\uparrow_{+2}&=&\bigg(\downarrow \downarrow\downarrow=-\frac 32\bigg)+(L_z=+2)
\eea
On the LF, the leading twist operators for the proton $uud$ with positive parity,
are typically of the form $$(u_\uparrow C\gamma^+u_\uparrow )d_\downarrow \qquad (u_\uparrow C i\sigma^{+i}u_\uparrow)d_\downarrow$$
with twist $\tau=\frac 72-\frac 32=2$, and $C=i\gamma^2\gamma^0$ the charge conjugation matrix.
Note that the two independent  Ioffe$^\prime$s currents
$$(u_\uparrow C\gamma_\mu u_\uparrow )\gamma^5\gamma^\mu d_\downarrow\qquad (u_\uparrow C i\sigma_{\mu\nu} u_\uparrow )\gamma^5\sigma^{\mu\nu} d_\downarrow$$
are twist $\tau=\frac 92-\frac 32=3$ on the LF, hence subleading.
In  terms of the good component quark fields $q_{+S}^C$, the
leading twist nucleon DAs for a spin-up proton are tied to the Fock states (\ref{eqn_Ji_WF}) and read~\cite{Belitsky:2005qn}
\begin{widetext}
\bea
\label{PDAs}
&&{\bf L_z=0:}\nonumber\\
&&\frac{\epsilon^{ABC}}{\sqrt{6}}\frac 1{p^+}\langle 0|(u^A_{+\uparrow}(z_1)C\gamma^+ u_{+\downarrow}^B(z_2))d^C_{+\uparrow}(z_3)|p^\uparrow\rangle
=\psi_1(z_1,z_2,z_3) N_{+}^\uparrow (p)\nonumber\\
&&\frac{\epsilon^{ABC}}{\sqrt{6}}\frac 1{p^+}\langle 0|(u^A_{+\uparrow}(z_1)C i\sigma^{+i} u_{+\uparrow}^B(z_2))d^C_{+\downarrow}(z_3)|p^\uparrow\rangle
=(\psi_2(z_1,z_3,z_2) +\psi_2(z_2,z_3,z_1) )\gamma^iN_{+}^\uparrow (p)
\\
\nonumber\\
&&{\bf L_z=+1:}\nonumber\\
&&\frac{\epsilon^{ABC}}{\sqrt{6}}\frac 1{p^+}\langle 0|(u^A_{+\uparrow}(z_1)C\gamma^+ u_{+\downarrow}^B(z_2))d^C_{+\downarrow}(z_3)|p^\uparrow\rangle
=(-i\slashed{\nabla}_{1\perp}\psi_3(z_1,z_2,z_3)-i\slashed{\nabla}_{2\perp}\psi_4(z_1,z_2,z_3))N_{+}^\uparrow (p)\nonumber\\
&&\frac{\epsilon^{ABC}}{\sqrt{6}}\frac 1{p^+}\langle 0|(u^A_{+\downarrow}(z_1) C i\sigma^{+i} u_{+\downarrow}^B(z_2))d^C_{+\uparrow}(z_3)|p^\uparrow\rangle
\nonumber\\
&&\qquad\qquad\qquad\qquad\qquad\qquad=\bigg(i\nabla_{+1}^i(\psi_4(z_3,z_1,z_2) -\psi_3(z_3,z_1,z_2)-\psi_3(z_3,z_2,z_1) )\nonumber\\
&&\qquad\qquad\qquad\qquad\qquad\qquad\qquad  +(i\nabla_{+2}^i(\psi_4(z_3,z_2,z_1) -\psi_3(z_3,z_2,z_1)-\psi_3(z_3,z_1,z_2) ))\bigg)
\gamma^iN_{+}^\uparrow (p)
\\
\nonumber\\
&&{\bf L_z=-1:}\nonumber\\
&&\frac{\epsilon^{ABC}}{\sqrt{6}}\frac 1{p^+}\langle 0|(u^A_{+\uparrow}(z_1) iC\sigma^{+i} u_{+\uparrow}^B(z_2))d^C_{+\uparrow}(z_3)|p^\uparrow\rangle
\nonumber\\
&&=\bigg(
i\nabla_{-1}^i(\psi_5(z_1,z_3,z_2) -\psi_5(z_1,z_2,z_3)) +
i\nabla_{-2}^i(\psi_5(z_2,z_3,z_1) -\psi_5(z_2,z_1,z_3) ) 
\bigg) N_{+}^\uparrow (p)
\\
\nonumber\\
&&{\bf L_z=+2:}\nonumber\\
&&\frac{\epsilon^{ABC}}{\sqrt{6}}\frac 1{p^+}\langle 0|(u^A_{+\downarrow}(z_1) C i\sigma^{+i} u_{+\downarrow}^B(z_2))d^C_{+\downarrow}(z_3)|p^\uparrow\rangle
\nonumber\\
&&=\bigg(
i\nabla_{+1}^{\{i} i\nabla_{\perp 2}^{j\}}(\psi_6(z_1,z_3,z_2) +\psi_6(z_2,z_1,z_3)) -\psi_6(z_1,z_3,z_2)) -\psi_6(z_2,z_3,z_1)) \nonumber\\
&&+i\nabla_{+1}^{\{i} i\nabla_{\perp 1}^{j\}}(\psi_6(z_1,z_3,z_2) +i\nabla_{+2}^{\{i} i\nabla_{\perp 2}^{j\}}(\psi_6(z_2,z_1,z_3) 
\bigg) N_{+}^\uparrow (p)
\eea
\end{widetext}
Here $N_+^\uparrow (p)$ is the good nucleon on-shell component with $p^2=m_N^2$,  and the shorhand notation
for the tracelss symmetrization
\bea
\nabla^{\{i} \nabla^{j\}}=\nabla^i\nabla^j+\nabla^j\nabla^i-\delta^{ij}\nabla^k\nabla^k\nonumber\\
\eea

\section{Summary and Outlook}
This paper is in many respects a methodical paper, devoted to novel technical tools of few-body quantum mechanics. We now provide a brief summary, and put forth few outlooks.

 {\bf Summary:} We started this paper by clarifying the use of the permutation group $ S_3$, in particular its pertinent representations,
 essential for defining the symmetry properties of the wave functions for three quarks. As it is well known since the early 1960's,  the baryon spin and flavor (isospin) parts of the wavefunction for the S-shell, 
do not factorize, as only their permutation symmetric and non-factorizable combination  is allowed.

For the  P-shell states (negative parity baryons) we have in addition the orbital $L=1$ part of the wavefunctions to symmetrize. This is best achieved using the  analogues of the $\rho$-type or $\lambda$-type Jacobi coordinates. In this case we need to construct triple tensor products
of permutation matrices, and find their {\em totally symmetric} states. How to do that, without guessing and a loss of generality, is shown in Appendix A. A nontrivial result (not new but explicitly derived) is that there is a $unique$ permutation symmetric wave function. In the next D-shell, the coordinates appear as products of two, so one has to find the  totally symmetric wavefunction
of four objects, and so on. All of those are found by the proposed method.

Our other methodical suggestion is to use the natural spin-tensor notations for the wavefunctions, and apply symbolic manipulation
capabilities of programs like Maple or
Mathematica.
The use of symbolic ``monoms" is traditional, yet it is better done in generic basis.  Specifically, all possible  spin-isospin
``monoms' of three quarks are $4^3=64$. While for the nucleon only 9 monoms are needed,  for the P-shell and D-shell baryons, the
wavefunctions are much more involved. The universal spin-tensor  notations allows for  any type of symbolic operation when coded, e.g.
spin-orbit with differentiation etc. In Mathematica, whether an operator is acting on the wavefunction with 1 or 64 components,
makes no practical difference. 
     
As a demonstration of this  technique, we repeated the well known Isgur-Karl calculations of the  splitting of the five negative parity $N^*$ resonances. Unlike them, we explicitly construc the
wavefunctions with proper permutation symmetry, and not as certain limits of less complicated $\Sigma,\Lambda$ wavefunctions when
$m_s\rightarrow m_u,m_d$. Like them, we also show that using only the spin-spin and tensor forces (but without spin-orbit) we can get a very good
description of the mass spectrum for these states.

In this paper we stated that, apart from the  orbital part, the basic wavefunction of all $N^*$ should be spherically symmetric in 6 dimensions, $\phi(\sqrt{\vec \rho^2+\vec \lambda^2})$, yet we have not evaluated it. However, we note that in our recent study~\cite{Miesch:2023hjt}, we have used the appropriate ``hyperdistance approximation" 
and the reduced radial Schrodinger equation, to obtain such wavefunctions, for (flavor-symmetric) baryons and 
(all-charmed) tetraquarks. The reason we focused on all-charm hadrons first, is that 
for heavy quarks the spin-dependent interactions -- coming from relativistic corrections -- are small and can be neglected in the zeroth approximation. 

Construction of LF wave functions for negative parity baryons needs explicit definition  of the longitudinal wave functions
(depending on momentum fractions $x_1,x_2,x_3$) with symmetries
of Jacobi coordinates $\rho,\lambda$. We have shown what these
functions are in section \ref{sec_LF_long}.

{\bf Outlook:} All of the present study is a preparation leading to the proper definitions of the the Light Front wave functions (LFWFs) for multiquark hadrons, or their multiquark components. What we mean can be explained by examples: to a heavy $\bar Q Q$ mesons,  
we can add a light $\bar q q$  pair to form a  $tetraquark$, while to a baryon we can
add a pair to turn it to a $pentaquark$. With pertinent  quantum numbers, these are ``exotic hadrons", minimizing the Hamiltonian. Also, they can be considered as   ``virtual clouds" (as is the 5-quark component of the nucleon, seeding its antiquark sea, 
e.g. discussed in \cite{Shuryak:2022wtk}
)
or separate ``pentaquark states" orthogonal to the nucleon with all its cloud (as 
the recently discovered pentaquark resonances with hidden charm $uud+c\bar c$ states). 
Their discovery was helped by their  narrow widths, as they happen to be just above the  thresholds of baryon-meson states to which they can decay. Te further discoveries of
stable pentaquarks below such thresholds (e.g. with  hidden $\bar b b$) are still ahead of us. Those with $\bar u u, \bar d d, \bar s s$ are perhaps not likely to be seen as separate resonances. However, the theory
still needs to answer:  where are they? what contribution to the known states (e.g. nucleon) they actually make?

When there are several identical quarks, the issues of  
 appropriate symmetries of the wavefunctions need to be resolved.  
 We have discussed in details the  S-shell and P-shell baryons $N^*$. Our important 
point was that the LFWFs should have {\em the same structure under the $S_3$ permutation group}  for their spin-isospin wavefunctions, as they have in the  CM approach.
The orbital part should not only contain the transverse $k_\perp^\pm$ momenta, as was proposed, but also it should include the longitudinal doubly degenerate excited states. We showed that those are
$D^{\rho,\Lambda}_{n_L,m_L}$  consisting of six waves on an equilateral triangle for longitudinal momenta fractions. 
We have further shown, that while spherical symmetry is strictly speaking absent on the LF, the numerical deviations
from the energies (and fixed-$J$ wave functions in the CM frame) for P-shell and D-shell states constitute 
only several percents. Hopefully these differences are smaller than the splitting generated by the spin-spin and tensor forces, In this way,
the squared mass splittings will not be affected by these ``non-sphericity" corrections. 

As parting comments, we note that
one would not be able to calculate the $\bar u,\bar d, \bar s, \bar c$ contributions to the PDFs and other density matrices, $without$ solving the many-body Hamiltonian, and
obtain the full wavefunctions. And, e.g. for the
five-quark components of the LFWFs with flavors content $uudu\bar u,uudd\bar d $, 
one would need also to combine the isospin representations  with the color,spin and orbital wave functions, to get the correct Fermi statistics. We hope to address those issues
in subsequent publications.
\\
\\
\\
{\bf Acknowledgements:}
\\
This work  is supported by the Office of Science, U.S. Department of Energy under Contract No. DE-FG-88ER40388.

\appendix
\section{Permutations and the wave functions} \label{sec_permutations}
\subsection{Permutation group $S_3$}

The (12) and (23) permutations are realized as improper O(3) rotations on the Jacobi 2-vector,
with determinant equal to $-1$.
This is rather natural if we think of the initial three $1,2,3$ particles set on an equilateral triangle.
The six permutations (\ref{PERM}) are then 3 in-plane rotations by $\frac \pi 3$ of which $P_4$ is an example, and three 
out-of plane rotations by $\pi$ along each of the 3 bi-sectors of which $P_2$ is an example.

 With this in mind, 
the Jacobi 2-vector under the 3-particle permutation, transforms analogously to a spin-$\frac 12$ 
under SU(2) rotations. In retrospect this is expected, since the mixed representation Young tableau
for 3 spin-$\frac 12$ carries the same dimension $2_M$ as the primitive spin-$\frac 12$ Young tableau.
This observation will be repeatedly used below to construct the excited states of baryons with 3 quarks,
with totally symmetric space-spin-flavor wavefunctions under $S_3$.

In general, the representations of  $S_3$ fall into: 1/ totally symmetric states ($S$), totally antisymmetric states ($A$),
and states with mixed symmetry ($M$), under $P_i$.  
Three quark spin states can be split into $S=\frac 32$ symmetric states, $\uparrow \uparrow \uparrow $ and so on,
and then two mutually orthogonal $S=\frac 12$ states which, following Jacobi coordinates, we call $S^\rho,S^\lambda$
states. The following development consists of two steps. The first one, going back at least to  \cite{Isgur:1978xj},
is related to symmetry under $(12)$ reflection (\ref{eqn_perm_12}).
With this in mind, any characterization of a 3-particle state  (space, spin, flavor) can be composed by paralleling
the Jacobi coordinates, with manifest symmetry under $S_3$.  
The idea,  is to construct mixed symmetry wave functions
 $M^{\rho, \lambda}$, out of $\rho$-like and $\lambda$-like blocks which have pure S or A permutation properties,  
\bea
\label{eqn_perm_M12}
[P_2=(12)]
\begin{pmatrix}
M^\rho\\
M^\lambda
\end{pmatrix}
&=&
\begin{pmatrix}
-1&0\\
0 & 1
\end{pmatrix}
\begin{pmatrix}
M^\rho\\
M^\lambda
\end{pmatrix}\nonumber\\
{[P_4=(23)]}
\begin{pmatrix}
M^\rho\\
M^\lambda
\end{pmatrix}
&=&
\begin{pmatrix}
\frac 12&\frac{\sqrt 3}2\\
\frac{\sqrt{3}}2& -\frac 12
\end{pmatrix}
\begin{pmatrix}
M^\rho\\
M^\lambda
\end{pmatrix}
\eea

The construction of the representations of $S_3$ is carried as for any other groups, e.g. through the familiar
generalization from the spinor representations of $O(3)$ to spin-1, spin-$\frac 32$ etc. 
The tensor  product of two generic representations $X_a$ and $X_b$  with different symmetries under $S_3$,
is a sum of representations $X_{ab}$, each with symmetries $S, A, M^{\rho,\lambda}$. While it  is clear  that the symmetries of the $S,A$ products are
\bea
&&S_a\otimes S_b=S_{ab}\nonumber\\
&&A_a\otimes A_b=S_{ab}\nonumber\\
&&S_a\otimes A_b=A_{ab}
\eea
the product of the mixed representations $M^{\rho,\lambda}$ viewed as a primitive $2_M$ doublets, is more subtle
and requires more detailed studies.

For two $\rho,\lambda$-type blocks there are $2^2=4$ combinations, that can be constructed using  say
a basis $(X^\rho_1X^\rho_2,X^\lambda_1 X^\rho_2,  X^\rho_1X^\lambda_2 , X^\lambda_1X^ \lambda_2)$. Half of the states are symmetric and half are antisymmetric under  12-interchange. Yet what needs to be done,  is to enforce say  23-permutation on these combinations.  Using (\ref{eqn_perm_12}) two  
times, we obtain
 matrices corresponding to this permutation  for two objects.  The corresponding M(23) matrix in this basis takes the form
%\begin{widetext}
\ba \label{eqn_XX}
%M_{23}\bigotimes M_{23} = \\
\begin{bmatrix} 
1/4 & \sqrt{3} /4&  \sqrt{3} /4& 3/4 \\
 \sqrt{3}/4 &   3/4 & -1/4 & - \sqrt{3}/4 \\ 
\sqrt{3}/4& -1/4&  3/4 & - \sqrt{3}/4 \\
   3/4  & - \sqrt{3}/4 & - \sqrt{3}/4 & 1/4 \end{bmatrix}.
\ea
%\end{widetext}
The commutator of M(12) with M(23) matrices, yields two
imaginary and two zero eigenvalues. The latter correspond to the $totally$ symmetric and antisymmetric combinations, respectively. The symmetric in this notations is
 $(1 ,0, 0, 1)$ or  $X^\rho_1X^\rho_2+X^\lambda_1X^ \lambda_2$,  the only combination of two objects
   $simultaneously$ symmetric under (12) and (23) (and in fact all) permutations. 
 This  
  is the one which we are seeking for the wavefunctions of the baryons in the ground S-shell, as well as e.g. the three spin-$\frac 32$ nucleons in the P-shell. Although it may indeed be easy to simply guess this form, the method we follow eliminates any guessing, and can be used for any number of objects.

In analogy of what we do for the rotational (spin) group $O(3)$, we derived the tensor product of two
representations into irreducible representations of $S_3$  
$$2_M\otimes 2_M=1_A\oplus 3_S$$
with $1_A$ a singlet anti-symmetric, and $3_{S}$ a triplet of symmetric representations.
The singlet $1_A$ antisymmetric representation is 
\bea
A_{ab}=\frac 1{\sqrt 2} M_a^T
\begin{pmatrix}
0&1\\
-1 & 0
\end{pmatrix}
M_b=\frac 1{\sqrt 2}(M_a^\rho M_b^\lambda-M_a^\lambda M_b^\rho)\nonumber\\
\eea
%which is readily checked to be invariant under all the six rotations. 
while one of the triplet $3_S$  is e.g.
\bea
M_{ab}^\rho=\frac 1{\sqrt 2} M_a^T
\begin{pmatrix}
0&1\\
1 & 0
\end{pmatrix}
M_b=\frac 1{\sqrt 2}(M_a^\rho M_b^\lambda+M_a^\lambda M_b^\rho)\nonumber\\
\eea
The remaining part of the triplet 
$3_S$ with projection-$\pm$, are regrouped in the manifestly orthogonal combinations
\bea
S_{ab}=\frac 1{\sqrt 2} M_a^T
\begin{pmatrix}
1&0\\
0 & 1
\end{pmatrix}
M_b=\frac 1{\sqrt 2}(M_a^\rho M_b^\rho+M_a^\lambda M_b^\lambda)\nonumber\\
\eea
\bea
M_{ab}^\lambda=\frac 1{\sqrt 2} M_a^T
\begin{pmatrix}
1&0\\
0 & -1
\end{pmatrix}
M_b=\frac 1{\sqrt 2}(M_a^\rho M_b^\rho-M_a^\lambda M_b^\lambda)\nonumber\\
\eea
$S_{ab}$ is invariant under all the six rotations, hence all the six permutations. It  is   manifestly symmetric.
The  combinations $M^{\rho,\lambda}_{ab}$ can be checked to transform as a doublet under all
permutations, e.g.
\bea
{[P_4=(23)]}
\begin{pmatrix}
M_{ab}^\rho\\
M_{ab}^\lambda
\end{pmatrix}
&=&
\begin{pmatrix}
\frac 12&\frac{\sqrt 3}2\\
\frac{\sqrt{3}}2& -\frac 12
\end{pmatrix}
\begin{pmatrix}
M_{ab}^\rho\\
M_{ab}^\lambda
\end{pmatrix}\nonumber\\
\eea
With this in mind, the spin-flavor combination
 \bea
S_{ab}= \frac 1{\sqrt 2}(S_{a}^\rho F_{b}^\rho+S_{a}^\lambda F_{b}^\lambda)
 \eea
 uniquely defines the symmetric part of the proton wave-function. 
  
 For three objects of $\rho,\lambda$ types, there are $2^3$ combinations possible, written e.g. in the basis of the following 8 monoms, 
 $$X^\lambda_1X^\lambda_2 X^\lambda_3, X^\rho_1X^\lambda_2 X^\lambda_3 , 
 X^ \lambda_1  X^\rho_2 X^\lambda_3, X^\rho_1   X^\rho_2 X^\lambda_3,  
 $$
$$  X^\lambda_1 X^\lambda_2 X^\rho_3, X^\rho_1 X^ \lambda_2  X^\rho_3, X^\lambda_1 X^\rho_2 X^\rho_3, X^\rho_1 X^\rho_2 X^\rho_3 $$
One should  proceed as we did before, calculating their transformation under (23) 
\begin{widetext}
\ba M_{23}^{3\,objects}
&=&\begin{bmatrix} 
-1/8 & \sqrt{3}/8& \sqrt{3}/8& -3/8& \sqrt{3}/8& -3/8&, -3/8,&  3 \sqrt{3}/8 \\
  \sqrt{3}/8,& 1/8,& -3/8,& -\sqrt{3}/8,& -3/8,& -\sqrt{3}/8,& 3 \sqrt{3})/8,& 3/ 8 \\ 
  \sqrt{3}/8,& -3/8,& 1/8,& -\sqrt{3}/8,& -3/8,& 3 \sqrt{3}/
  8,& -\sqrt{3}/8,& 3/8 \\ 
  -3/8,& -\sqrt{3}/8,& -\sqrt{3}/8,& -1/8,& 3 \sqrt{3}/8,& 3/8,&
   3/8,& \sqrt{3}/8 \\
   \sqrt{3}/8,& -3/8,& -3/8,& 3 \sqrt{3}/8,& 1/
  8,& -\sqrt{3}/8,& -\sqrt{3}/8,& 3/8 \\
   -3/8,& -\sqrt{3}/8,& 
  3 \sqrt{3}/8,& 3/8,& -\sqrt{3}/8,& -1/8,& 3/8,& \sqrt{3}/8 \\
  -3/8,& 3 \sqrt{3}/8,& -\sqrt{3}/8,& 3/8,& -\sqrt{3}/8,& 3/8,& -1/8,& \sqrt{3}/
  8 \\ 3 \sqrt{3}/8,& 3/8,& 3/8,& \sqrt{3}/8,& 3/8,& \sqrt{3}/8,& \sqrt{3}/8,& 
  1/8 
\end{bmatrix} \nonumber \ea
%1/8 & \sqrt{3}/8 & \sqrt{3}/8& 3/8& \sqrt{3}/8& 3/8& 3/8& (3 \sqrt{3})/
%  8 \\
%   -(\sqrt{3}/8)& 1/8& -(3/8)& \sqrt{3}/8& -(3/8)& \sqrt{3}/
%  8& -((3 \sqrt{3})/8)& 3/8}\\ 
%  -(\sqrt{3}/8)& -(3/8)& 1/8& \sqrt{3}/
%  8& -(3/8)& -((3 \sqrt{3})/8)& \sqrt{3}/8& 3/8 \\
%   3/8& -(\sqrt{3}/8)& -(\sqrt{3}/8)& 1/8& (3 \sqrt{3})/8& -(3/8)& -(3/8)& 
%  \sqrt{3}/8}\\
%  -(\sqrt{3}/8)& -(3/8)& -(3/8)& -((3 \sqrt{3})/8)& 1/8& 
%  \sqrt{3}/8& \sqrt{3}/8& 3/8}& \\ 3/8& -(\sqrt{3}/8)& (3 \sqrt{3})/
%  8& -(3/8)& -(\sqrt{3}/8)& 1/8& -(3/8)& \sqrt{3}/8} \\
%  3/8& (3 \sqrt{3})/
%  8& -(\sqrt{3}/8)& -(3/8)& -(\sqrt{3}/8)& -(3/8)& 1/8& \sqrt{3}/8}\\
%   -((3 \sqrt{3})/8)& 3/8& 3/8& -(\sqrt{3}/8)& 3/8& -(\sqrt{3}/8)& -(\sqrt{3}/8)& 1/8
\end{widetext}
Although this matrix may appear involved,   its determinant is 1, with four eigenvalues $(-1)$ and four $(+1)$.
Note that it is the same set of eigenvalues as for the (diagonal) matrix of the (12) permutation.
So, there is a 4-dimensional subspace which is symmetric under (23).
 The other 4-dimensional subspace (half of our basis) corresponds to the symmetric combinations under (12).
The eigensystem of the commutator of these matrices, yields
two zero eigenvalues, of which {\em only one}
is the symmetric combination, $(-1, 0, 0, 1, 0, 1, 1, 0)$ in our basis, corresponding to 
the sought after symmetric wavefunction under $S_3$,
\be \label{eqn_sym_in_3}
-X^\lambda X^\lambda X^\lambda +X^\rho X^\rho X^\lambda+ X^\rho X^\lambda X^\rho+X^\lambda X^\rho X^\rho \ee  Again, no guessing is needed,
the  $unique$   totally symmetric  wave functions is shown by construction.
We recall that while our  derivation is generic, we used it for three blocks here being the orbital, spin and isospin parts of the wavefunction.

This method can be used for any number of blocks, and we now proceed to four, as e.g. those needed for  the D-shell resonances.
The matrix for the  23-permutation in this case is $16\times16$, with 8 antisymmetric and 8 symmetric eigenvectors. In fact it can be generated in Mathematica by using command ${\bf KroneckerProduct[P23,P23,P23]}$ (in which case
the basis set is automatically selected by Kroneker product as well). The commutator of (12) and (23) matrices has 6 zero
eigenvalues, of which 3 are symmetric. Their linear combinations can be used
as wavefunctions in the D-shell. They are 
\ba \label{eqn_sym_in_4}
\{1, 0, 0, 0, 0, 0, 1, 0, 0, 1, 0, 0, 0, 0, 0, 1\}\nonumber \\
\{0, 0, 0, 1, 0, 0, -1, 0, 0, -1, 0, 0, 1, 0, 0, 0\}\nonumber \\
\{0, 0, 0, 0, 0, 1, -1, 0, 0, -1, 1, 0, 0, 0, 0, 0\}
\ea
Note that all of them are combination of four
``monoms", weighted by simple $\pm 1$ coefficients.

Although we do not have immediate applications for it, we repeated the procedure for 5 blocks, in the representation of dimension $2^5=32$. The corresponding (12) and (23) matrices   were generated. Their commutator has 10 zero eigenvalues, with half corresponding to symmetric and half antisymmetric wavefunctions. The corresponding
combinations are more complicated and have more than 4 nonzero terms.

\section{Details of matrix element calculations}
%\subsection{Splitting of states by spin-dependent forces}
%We have calculated matrix elements of all the forces discussed in two ways, analytically using symbolic
%form of the wave functions, and numerically, using spin-tensor form of those in Mathematica.
We have calculated the matrix elements of all the operators in two  ways:  by direct calculations using the
explicit wavefunctions quoted earlier, and  using symbolic manipulations of the wavefunctions represented as spin-tensors via Mathematica. We
will detail here some parts of the explicit calculations for all the spin interactions, starting from the unmixed states with $J=\frac 52$, then proceed to the mixing matrix in the $J=\frac 32$ shell, and only quote 
the final results for the mixing matrix in the $J=\frac 12$ shell.

\subsection{Spin-spin coupling in $J=\frac 32$}
Using the explicit states (\ref{SPIN32}-\ref{SPIN12}) and short-hand notations, the spin-spin coupling
contributions gives
\begin{widetext}
\bea
\label{3232}
\langle \frac 32|\mathbb V_{S}|\frac 32\rangle_{J=\frac 32}=
\frac 32\bigg((\varphi^\rho \mathbb V_S(\sqrt 2\rho)\varphi^\rho+\varphi^\lambda \mathbb V_S(\sqrt 2\rho)\varphi^\lambda)\,
(S_{\frac 32}^S\frac 14 \sigma_1\cdot\sigma_2 S_{\frac 32}^S)
%(\varphi^\rho  S^S_{\frac 32}\mathbb V_T\varphi^\rho S^S_{\frac 32})
\bigg)
\eea
\end{widetext}
with the Clebsches and some indices omitted to avoid clutter.
The overall factor of 3 follows from the permutation symmetry as we noted.
%The 12-part of the tensor $\mathbb V_T$ measures only the quadrupole deformation 
%in the $\rho$-coordinate, and therefore averages to zero in $\varphi^\lambda$.  
The spin-spin contribution is reduced by the identity
\bea
\label{SPIN1}
\sigma_1\cdot \sigma_2 S^S_{\frac 32, \frac 12}=\frac 16(4S_{\rm tot}^2-9)S^S_{\frac 32, \frac 12}=
S^S_{\frac 32, \frac 12}
\eea
in the permutation symmetric  S-state,  and gives
\bea
\label{3232S}
\langle \frac 32|\mathbb V_{S}|\frac 32\rangle_{J=\frac 32}= \frac 14 \int d\vec\rho\, d\vec \lambda\,({\vec\rho}^2+{\vec\lambda}^2)\, |\varphi_{00}|^2\,\mathbb V_S(\sqrt 2 \rho)\nonumber\\
\eea
The same arguments yield 
\bea
\label{3232SS}
\langle \frac 12|\mathbb V_{S}|\frac 12\rangle_{J=\frac 32}&=&
-\frac 14  \int d\vec\rho\, d\vec\lambda\, ({\vec\rho}^2+{\vec\lambda}^2)\, |\varphi_{00}|^2\,\mathbb V_S(\sqrt 2 \rho)\nonumber\\
\langle \frac 32|\mathbb V_{S}|\frac 12\rangle_{J=\frac 32}&=&0\nonumber\\
\eea
where we used the spin identities
\bea
\sigma_1\cdot\sigma_2S^\lambda_{\frac 12 \frac 12}&=&+S^\lambda_{\frac 12 \frac 12}\nonumber\\
\sigma_1\cdot\sigma_2S^\rho_{\frac 12 \frac 12}&=&-3S^\rho_{\frac 12 \frac 12}
\eea

\subsubsection{Tensor coupling in $J=\frac 32$}
The tensor contribution mixes configurations with different spin content, and is more involved. To evaluate it, we re-establish the Clebsches and azimuthal labelings
\begin{widetext}
\bea
\label{3232T}
\langle \frac 32|\mathbb V_{T}|\frac 32\rangle_{J=\frac 32}=\frac 32\bigg(
\bigg(\sqrt{\frac 35} \varphi^\rho_{10}  S^S_{\frac 32 \frac 32}-\sqrt{\frac 25}\varphi^\rho_{11}  S^S_{\frac 32 \frac 12}\bigg)
\mathbb V_T
\bigg(\sqrt{\frac 35} \varphi^\rho_{10}  S^S_{\frac 32 \frac 32}-\sqrt{\frac 25}\varphi^\rho_{11}  S^S_{\frac 32 \frac 12}\bigg)
\bigg)
\eea
\end{widetext}
%\bea
%\label{SPIN2}
%&&S_{\frac 32, \frac 12}^S\frac 12(\sigma_1^a\sigma_2^b+\sigma_1^b\sigma_2^a)S_{\frac 32, \frac 12}^S\nonumber\\
%&&=\frac 12 \delta^{ab}\,S_{\frac 32, \frac 12}^S\sigma_1\cdot \sigma_2 S_{\frac 32, \frac 12}^S=\frac 12 \delta^{ab}
%\eea
%which imply that the tensor contribution in (\ref{3232}) vanishes by contraction. Hence, (\ref{3232}) reduces to the single radial integral
%\bea
%\label{3232X}
%\langle \frac 32|\mathbb V_{S+T}|\frac 32\rangle=\frac 32 \int d\rho d\lambda (\lambda^2+\rho^2)\, |\varphi_{00}|^2\,\mathbb V_S(\sqrt 2 \rho)\nonumber\\
%\eea
where only the manifestly non-zero contributions are retained.
Using the spin-space dependence of the tensor interaction, we find that the spin valued parts in (\ref{3232T}) can be reduced, using the 
matrix elements
\bea
\label{3232T1}
S^S_{\frac 32 \frac 32}\mathbb V^\rho_T S^S_{\frac 32 \frac 32}&=& +\frac 23\sqrt{\frac \pi 5}\,Y_2^0(\hat \rho)\nonumber\\
S^S_{\frac 32 \frac 12}\mathbb V^\rho_T S^S_{\frac 32 \frac 12}&=&-\frac 23\sqrt{\frac \pi 5}\, Y_2^0(\hat \rho)\nonumber\\
S^S_{\frac 32 \frac  12}\mathbb V^\rho_T S^S_{\frac 32 \frac 32}&=& -\frac 23\sqrt{\frac{2 \pi }5}\,Y_2^{+1}(\hat \rho)\nonumber\\
S^S_{\frac 32 \frac  32}\mathbb V^\rho_T S^S_{\frac 32 \frac 12}&=& +\frac 23\sqrt{\frac{2 \pi} 5}\,Y_2^{-1}(\hat \rho)
\eea
with
\bea
\mathbb V^\rho_T = \frac 12\bigg(\sigma_1\cdot\hat\rho\,\sigma_2\cdot\hat\rho-\frac 13 \sigma_1\cdot\sigma_2\bigg)
\eea
\begin{widetext}
to give in short hand notations
\bea
\label{3232T2}
\langle \frac 32|\mathbb V_{T}|\frac 32\rangle_{J=\frac 32}=\frac{\sqrt\pi}{\sqrt 5}
\bigg(\frac 35 \varphi_{10}Y_2^0\varphi_{10}-\frac 25 \varphi_{11}Y_2^0\varphi_{11}
-\frac {2\sqrt{3}}5\varphi_{10}Y_2^{-1}\varphi_{11}+\frac {2\sqrt{3}}5\varphi_{11}Y_2^{+1}\varphi_{10}\bigg)
\eea
\end{widetext}
If we recall that
\bea
\varphi_{1m}^\rho=\sqrt{\frac {8\pi}3}\,Y_1^m(\hat \rho)\,\rho\varphi_{00}\nonumber\\
\varphi_{1m}^\lambda=\sqrt{\frac {8\pi}3}\,Y_1^m(\hat \lambda)\,\lambda\varphi_{00}
\eea
the integration over the three spherical harmonics in (\ref{3232T2}) can be undone using the identity
\begin{widetext}
\bea
\int d\hat \rho\, Y_{l_1}^{m_1}(\hat \rho)\, Y_{l_2}^{m_2}(\hat \rho)\, Y_{l_3}^{m_3}(\hat \rho)=
\bigg(\frac{(2l_1+1)(2l_2+1)(2l_3+1)}{4\pi}\bigg)^{\frac 12}
\begin{pmatrix}
l_1 & l_2 & l_3 \\
 0 & 0 & 0
 \end{pmatrix}
 \begin{pmatrix}
l_1 & l_2 & l_3 \\
m_1 & m_2 &m_3
 \end{pmatrix}
 \eea
\end{widetext}
with the result
\bea
\label{3232T}
\langle \frac 32|\mathbb V_{T}|\frac 32\rangle_{J=\frac 32}=
\frac {4}{15}\int d\vec\rho\, d\vec\lambda\,\vec\rho^2|\varphi_{00}|^2\,\mathbb V_T(\sqrt 2\rho)\nonumber\\
\eea
The remaining spin-spin and tensor matrix elements can be done similarly, with the results 
\begin{widetext}
\bea
\label{3232T}
\langle \frac 32|\mathbb V_{T}|\frac 12\rangle_{J=\frac 32}=
-\frac 1{3\sqrt{10}} \int d\vec\rho\, d\vec\lambda\,\vec\rho^2|\varphi_{00}|^2\,\mathbb V_T(\sqrt 2\rho)
\qquad\qquad  \langle \frac 12|\mathbb V_{T}|\frac 12\rangle_{J=\frac 32}=0
\eea
\end{widetext}
for the remaining mixing matrix entries in the $J=\frac 32$ P-shell. We used the fact that the off-diagonal and non-vanishing tensor matrix elements are
\bea
S^S_{\frac 32 \frac 32}\mathbb V^\rho_T S^\lambda_{\frac 12 \frac 12}&=& +\frac 23\sqrt{\frac \pi 5}\,Y_2^{-1}(\hat \rho)\nonumber\\
S^S_{\frac 32 \frac 12}\mathbb V^\rho_T S^\lambda_{\frac 12 \frac 12}&=&-\frac 23\sqrt{\frac {2\pi} 5}\, Y_2^0(\hat \rho)
\eea
with the vanishing diagonal ones
\bea
S^{\rho, \lambda}_{\frac 12 \frac 12}\mathbb V^\rho_T S^{\rho,\lambda}_{\frac 12 \frac 12}=0
\eea

\subsubsection{Spin-orbit coupling in $J=\frac 32$}

 When we calculate  the $LS$ interaction of one pair of quarks, 1-2, all
  orbital terms  $\sim \lambda$   are eliminated by derivative over $\vec\rho$  and only $\rho$-dependent ones remain.
  
The spin-orbit contributions simplify considerably, if we note that $\varphi^\rho$ spins along the $\lambda$-direction,
and $\varphi^\lambda$ spins along the $\rho$-direction. This means that the standard spin-orbit  contribution in (\ref{SL1})
has only a non-vanishing matrix element for the combination
\bea
\varphi^\rho (\rho\times p_\rho)\varphi^\rho=l_\rho 
\eea
as the two other entries vanish 
\bea
\varphi^\lambda (\rho\times p_\rho)\varphi^\rho=\varphi^\lambda (\rho\times p_\rho)\varphi^\lambda=0
\eea
The mixed spin-orbit contribution in (\ref{SL1}) vanishes identically, as all entries
\bea
\varphi^{\rho,\lambda}(\rho\times p_\lambda)\varphi^{\rho, \lambda}=0
\eea
are seen to angle average to zero. As a result, the 12-spin-orbit contribution (\ref{SL1}) simplifies to
\bea
\label{SL1X}
\mathbb V_{SL}(1,2)\rightarrow V_{SL}(\sqrt 2\rho)\,\frac 12
(\sigma_1+\sigma_2)\cdot l^\rho
\eea

The  spin-orbit contribution mixes configurations with different spin content as well. Its evaluation follows that of the 
tensor coupling detailed  above. More specifically, we have
\begin{widetext}
\bea
\label{3232SL}
\langle \frac 32|\mathbb V_{SL}|\frac 32\rangle_{J=\frac 32}=\frac 32\bigg(
\bigg(\sqrt{\frac 35} \varphi^\rho_{10}  S^S_{\frac 32 \frac 32}-\sqrt{\frac 25}\varphi^\rho_{11}  S^S_{\frac 32 \frac 12}\bigg)
\mathbb V_{SL}
\bigg(\sqrt{\frac 35} \varphi^\rho_{10}  S^S_{\frac 32 \frac 32}-\sqrt{\frac 25}\varphi^\rho_{11}  S^S_{\frac 32 \frac 12}\bigg)
\bigg)
\eea
%\end{widetext}
We can simplify the orbital contributions in (\ref{3232SL}) if we recall that
\bea
l_\rho^z\varphi^\rho_{1m}=m\varphi^\rho_{1m}\qquad
l^\pm _\rho\varphi^\rho_{1m}=\sqrt{2-m(m\pm 1)}\varphi^\rho_{1m\pm1}\nonumber\\
\eea
with the result
\bea
\label{3232SL1}
\langle \frac 32|\mathbb V_{SL}|\frac 32\rangle_{J=\frac 32}=\frac 3{10}\bigg(2|\varphi^\rho_{11}|^2
S^S_{\frac 32 \frac 12}\frac 12 (\sigma_1+\sigma_2)^zS^S_{\frac 32 \frac 12}
-\sqrt{3}|\varphi^\rho_{10}|^2 \bigg( S^S_{\frac 32 \frac 32}\frac 12(\sigma_1+\sigma_2)^+S^S_{\frac 32 \frac 12} +{\rm c.c.}\bigg)
\bigg)
\eea
\end{widetext}
Using the permutation symmetry of the in-out spin symmetric S-states, we can make use of the identities
\bea
S^S_{\frac 32 \frac 12}\frac 12 (\sigma_1+\sigma_2)^zS^S_{\frac 32 \frac 12}&=&
\frac 23 S^S_{\frac 32 \frac 12}\,S^z_{\rm tot}\,S^S_{\frac 32 \frac 12}=\frac 13\nonumber\\
S^S_{\frac 32 \frac 32}\frac 12 (\sigma_1+\sigma_2)^+S^S_{\frac 32 \frac 12}&=&
\frac 23 S^S_{\frac 32 \frac 12}\,S^+_{\rm tot}\,S^S_{\frac 32 \frac 12}=\frac 2{\sqrt 3}\nonumber\\
\eea
which allow for the simplification of (\ref{3232SL1}) into
\bea
\label{3232SL2}
&&\langle \frac 32|\mathbb V_{SL}|\frac 32\rangle_{J=\frac 32}\nonumber\\
&&=\frac 15\int d\vec\rho\, d\vec\lambda\,(|\rho_-|^2-12\rho_z^2)\,|\varphi_{00}|^2\,\mathbb V_{SL}(\sqrt 2\rho)\nonumber\\
&&=-\frac 23 \int d\vec\rho\, d\vec\lambda\,\vec\rho^2\,|\varphi_{00}|^2\,\mathbb V_{SL}(\sqrt 2\rho)
\eea
The remaining spin-orbit contributions to the mixing matrix for $J=\frac 32$, can be found similarly with the results
\bea
\label{3232SL3}
\langle \frac 32|\mathbb V_{SL}|\frac 12\rangle_{J=\frac 32}&=&-\frac{\sqrt{10}}6
\int d\vec\rho\, d\vec\lambda\,\vec\rho^2\,|\varphi_{00}|^2\,\mathbb V_{SL}(\sqrt 2\rho
\nonumber\\
\langle\frac  12|\mathbb V_{SL}|\frac 12\rangle_{J=\frac 32}
%&&=2\int d\vec\rho \,d\vec\lambda\,|\rho_-|^2\,|\varphi_{00}|^2\,\mathbb V_{SL}(\sqrt 2\rho)\nonumber\\
&=&+\frac 13 \int d\vec\rho\, d\vec\lambda\,\vec\rho^2\,|\varphi_{00}|^2\,\mathbb V_{SL}(\sqrt 2\rho)\nonumber\\
\eea

\subsubsection{${}^\prime$t Hooft coupling in $J=\frac 32$}

The $^\prime$t Hooft coupling contribution to the mixing matrix, involves flavor matrix elements.
The calculations simplifies considerably if we note that the flavor states $F^{S,\rho, \lambda}$ 
are eigenstates of the flavor singlet projector,
\bea
(1-\tau_1\cdot \tau_2)\,F^S&=&0\,F^S\nonumber\\
(1-\tau_1\cdot \tau_2)\,F^\rho&=&4\,F^\rho\nonumber\\
(1-\tau_1\cdot \tau_2)\,F^\lambda&=&0\,F^\lambda\nonumber\\
\eea
With this in mind, a rerun of the preceding arguments gives
\begin{widetext}
\bea
\label{VTHOOFTX}
\langle \frac 32|\mathbb V_{TH}|\frac 32\rangle_{J=\frac 32}&=& \frac 85 (1-a)\,
\int d\vec\rho\, d\vec\lambda\,\vec\rho^2\,|\varphi_{00}|^2\,\mathbb V_{TH}(\sqrt 2\rho)\nonumber\\
\langle \frac 12|\mathbb V_{TH}|\frac 12\rangle_{J=\frac 32}&=& =2 \,
\int d\vec\rho\, d\vec\lambda\,
\bigg((1-a)\,\vec{\rho}^2+(1+3a)\,\vec{\lambda}^2\bigg)
\,|\varphi_{00}|^2\,\mathbb V_{TH}(\sqrt 2\rho)\nonumber\\
\eea
%\end{widetext}

\subsubsection{Spin splittings in the $J=\frac 52$ shell}
The spin $J=\frac 52$ shell is unmixed. The spin interactions in this shell are the simplest to evaluate. Using their explicit wavefunction, 
and some of the spin and flavor identities we derived earlier, we obtain
\bea
\langle \frac 32|\mathbb V_S|\frac 32\rangle_{\frac 52}&=&\frac 14 
\int d\vec\rho\,d\vec\lambda\,(\vec{\rho}^2+\vec{\lambda}^2)\,|\varphi_{00}|^2\,\mathbb V_S(\sqrt{2}\rho)\nonumber\\
\langle \frac 32|\mathbb V_T|\frac 32\rangle_{\frac 52}&=&-\frac 1{15}
\int d\vec\rho\,d\vec\lambda\,\vec{\rho}^2\,|\varphi_{00}|^2\,\mathbb V_T(\sqrt{2}\rho)\nonumber\\
\langle \frac 32|\mathbb V_{SL}|\frac 32\rangle_{\frac 52}&=&
\int d\vec\rho\,d\vec\lambda\,\vec{\rho}^2\,|\varphi_{00}|^2\,\mathbb V_{SL}(\sqrt{2}\rho)\nonumber\\
\langle \frac 32|\mathbb V_{TH}|\frac 32\rangle_{\frac 52}&=&4 (1-a)
\int d\vec\rho\,d\vec\lambda\,\vec{\rho}^2\,|\varphi_{00}|^2\,\mathbb V_{TH}(\sqrt{2}\rho)
\eea
with the short-hand notation  $|\frac 32\rangle_{\frac 52}\equiv |1\frac 32\frac 52\frac 52\rangle_{p^-}$.

\subsection{Mixing matrix}
The hyperfine interactions in the degenerate P-sub-shells of fixed $J=\frac 12, \frac 32$ are fixed by
\bea
\label{DEG1}
\mathbb M_J=\frac 1{\langle p^-|p^-\rangle}
\begin{pmatrix}
\langle \frac 32|\mathbb V_{S+T+SL+V_{TH}}|\frac 32\rangle  & \langle \frac 32|\mathbb V_{S+T+SL+V_{TH}}|\frac 12\rangle \\
\langle \frac 12|\mathbb V_{S+T+SL+V_{TH}}|\frac 32\rangle  & \langle \frac 12|\mathbb V_{S+T+SL+V_{TH}}|\frac12\rangle 
\end{pmatrix}_J
\eea
where we used the  S-labeling as a short-hand for the degenerate  nucleon P-states, which are normalized by
\bea
\label{PP}
\langle p^-|p^-\rangle=\frac 13 {\mathbb N_P}=\frac 13\int d\vec \rho\, d\vec\lambda \,(\vec\rho^2+\vec\lambda^2)\,|\varphi_{00}|^2
\eea
The diagonalization of (\ref{DEG1}) yields
the two mixing angles in the P-shell with $J=\frac 12, \frac 32$. For $J=\frac 32$, their explicit forms we already given in
results (\ref{eqn_results})

\end{widetext}
\section{Explicit wave functions of baryons in spin-tensor notations in Mathematica}

Mathematica is a platform for symbolic  calculations widely used in many branches
of physics. In some (e.g. general relativity) its ability to handle multi-component
tensors and perform e.g. hundreds of differentiations, is crucial for progress in 
the field, as recognized long ago. (A side remark about Maple:
of course one can do anything in it as well. Yet its elaborate structures --
{\bf sets, arrays, vectors, matrices} seem a bit cumbersome,  at least for 
beginners. A single notion of {\bf Table} in Mathematica is the only one needed.)

In quantum mechanics, with applications like the ones discussed in this paper, 
Matematica's ability to do $analytic$ (rather than numeric) computations
are not yet sufficiently utilized. 
A wavefunction may have hundreds or thousands of
components, which 
  can be naturally added, combined by tensor products,
acting upon by spin-dependent or differential operators, squared and integrated.  The rules of operation with Tables of any number 
of dimensions are simple or even elementary, yet some explanation may
be helpful to some readers. That is why in this paper we focused
on the 3-quark systems, as  their wavefunctions have $2^6=64$ ``monoms".
It is already complicated enough, so that issues of e.g. symmetries
of identical quarks are nontrivial. This way of solving the few-quark problems, should prove   useful for the
other (newly
discovered) tetraquarks,  and pentaquarks  states as well.

The usual representation of quantum states is done with certain {\em monoms} times cooordinate functions. The important case of  the nucleon is a well known example. Three light quarks, with 2 spin times 2 isospin
states, lives in a basis of $2^6=64$ possible monoms. Only nine are  in  (\ref{eqn_N_monoms})  for this example. A 
different number with other functional coefficients appear for the excited (e.g. P-shell) states.  The natural standardized way to
represent all wavefunctions is provided by {\em spin-tensors}. The suggested order of the indices in their definition
is arbitrary, so we propose to keep the spin and then isospin  $s_1,s_2,s_3,i_1,i_2,i_3$ sequentially, each with binary values 1 or 2. 
Up and down notations can be held either symbolically (in formulae) or explicitly, for which we define the  elementary states
$${\rm up}=\{1,0\}; {\rm down}=\{0,1\} $$
and then use the ${\bf substitution}$ function while forming actual spin-tensor. Let us take as an example the $S=\frac 12,S_z=\frac 12$ expressions
for the $\rho,\lambda$ blocks we used above
$ S^\rho= (\uparrow \downarrow \uparrow-\downarrow \uparrow \uparrow)/\sqrt{2} $
%S^\lambda&=& (\uparrow \downarrow \uparrow+\downarrow \uparrow \uparrow-2 \uparrow  \uparrow \downarrow)/\sqrt{6} \nonumber \ea
and show the Mathematica command converting it to explicit spin-tensor form
\begin{widetext}
{\footnotesize 
\ba S\rho&:=&(u[[s1]] d[[s2]]  u[[s3]] - d[[s1]]  u[[s2]]  u[[s3]] )/\sqrt{2}; \nonumber \\
S\rho^{numeric}&=&Table[(S\rho/. \{u -> up, d -> down\}, \{s1, 1, 2\}, \{s2, 1, 2\}, \{s3, 1, 2\}] \nonumber \\
 &=&\{\{\{0, 0\}, \{-1/\sqrt{2}, 0\}\}, \{\{1/\sqrt{2}, 0\}, \{0, 0\}\}\} \ea
 }
Note that we are using the $u,d$ notations in any symbolic expression. The explicit numerical form follows through the substitution to the $up,down$ monoms. 
While keeping unnecessary components represented by 6 zeros, it is still quite compact and convenient to use. 
The isospin expressions with $I=\frac 12$ are the same,  with only the indices redefined as $i1,i2,i3$. 

The spin-isospin wavefunction of the proton (spin up) is then 
{\footnotesize 
\ba 
p \uparrow &=& 
Table[ (S\rho[[s1, s2, s3]] S\rho[[i1, i2, i3]] + 
    S\lambda[[s1, s2, s3]]S\lambda[[i1, i2, i3]])/\sqrt{2} \nonumber \\
  &,&  \{s1, 1, 
    2\}, \{s2, 1, 2\}, \{s3, 1, 2\}, \{i1, 1, 2\}, \{i2, 1, 2\}, \{i3, 1, 2\}]  \ea
$$=  \{\{\{\{\{\{0, 0\}, \{0, 0\}\}, \{\{0, 0\}, \{0, 0\}\}\}, \{\{\{0, \sqrt{2}/
      3\}, 
      \{-1/(3 \sqrt{2}), 0\}\}, \{\{-1/(3 \sqrt{2}), 0\}, $$
$$      
      \{0, 
      0\}\}\}\}, \{\{\{\{0, -1/(3 \sqrt{2})\}, \{\sqrt{2}/3, 
      0\}\}, \{\{-1/(3 \sqrt{2}), 0\}, \{0, 0\}\}\}, \{\{\{0, 0\}, \{0, 0\}\}, $$
$$      \{\{0, 
      0\}, \{0, 0\}\}\}\}\}, \{\{\{\{\{0, -1/(3 \sqrt{2})\}, \{-1/(3 \sqrt{2}), 
      0\}\}, \{\{\sqrt{2}/3, 0\}, \{0, 0\}\}\}, \{\{\{0, 0\}, \{0, 0\}\}, \{\{0, 0\}, $$
$$      \{0, 
      0\}\}\}\}, \{\{\{\{0, 0\}, \{0, 0\}\}, \{\{0, 0\}, \{0, 0\}\}\}, \{\{\{0, 0\}, \{0, 
      0\}\}, \{\{0, 0\}, \{0, 0\}\}\}\}\} $$
      }
      In this example only 9 elements (out of 64) are nonzero. Te matrix elements are evaluated as usual, e.g. the
normalization is the sum over all indices
\ba \langle p | p \rangle = 
 Sum[p[[s1, s2, s3, i1, i2, i3]]^2, \{s1, 1, 2\}, \{s2, 1, 2\}, \{s3, 1, 
   2\}, \{i1, 1, 2\}, \{i2, 1, 2\}, \{i3, 1, 2\}]  \nonumber \ea
    If complex,  the outgoing wave function needs to be conjugated, as usual.
\end{widetext}
 
   For nonzero orbital momentum (e.g. $L=1$ to be discussed), the explicit orbital functions are linear in coordinates, and will  be defined below. 
  With 6 coordinates $\vec \rho, \vec \lambda$ in any matrix elements, we will 
 perform explicitly the integration over the 4 angles in  both solid angles,   $d\Omega_\rho d\Omega_\lambda$.  All  the wavefunctions to be shown below have 
 the same normalization factor 
\be N_{norm}={32 \pi^2 \over 3}\langle \lambda^2 + \rho^2 \rangle 
\ee
 by which the matrix elements of operators should be divided. (The meaning of
  angular brackets here was defined in (\ref{eqn_averaging}, it contains a double integral over the $\rho,\lambda$ moduli, with the remaining radial wave functions.).
  
  The matrix elements of any operators including spin, isospin and coordinate variables (in matrix or differential form) can easily be calculated from these wavefunctions, with the summation over all indices. For P-shell nucleons,  we
  constructed nine of them, with $|J,J_z,S>$ selections
  \ba && |{5 \over 2},{5 \over 2},{3 \over 2}> |{5 \over 2},{3 \over 2},{3 \over 2}>,|{5 \over 2},{1 \over 2},{3 \over 2}>, \nonumber \\
 && |{3 \over 2},{3 \over 2},{3 \over 2}>, |{3 \over 2},{1 \over 2},{3 \over 2}>, \nonumber \\
  &&  |{3 \over 2},{3 \over 2},{1 \over 2}>,|{3 \over 2},{1 \over 2},{1 \over 2}>,|
  \nonumber \\
 && |{1 \over 2},{1 \over 2},{3 \over 2}>,|{1 \over 2},{1 \over 2},{1 \over 2}> \nonumber \ea
After checking normalization and mutual orthogonality, the
  matrix elements of all spin-dependent interactions have been calculated. Since operators are rotationally scalars, all
  states which differ in $J_z$ (orientation) only, must yield the same matrix elements. 
  
  We explicitly show here five wave functions, all with $J_z=1/2$, for which
  pertinent Clebsch-Gordon coefficients and coordinates are included.  (Let us remind the reader  that those are shown $before$ mixing, when they still have
  definite total spins, $S=\frac 32$ and $S=\frac 12$.) We present those spin-tensors in
  full, with 6 indices and 6 curly brackets, and also keep all indices in operators
  to avoid confusion. 
  
  (While it is not necessary, one can further use even more compressed notations.
  By {\bf Flatten} command,  the wave functions are reduced from spin-tensors to 
  64-d vectors in the ``monom space".  All operators can also be redefined in it, as $64\times 64$ matrices.
  The benefit of this, is that in this form they can be multiplied as ordinary matrices. Let us demonstrate how it works for the spin-spin interaction\\
 \begin{widetext}
  \begin{verbatim} s := Table[1/2*PauliMatrix[i], {i, 1, 3}] 
  Slist[i_, m_] := Insert[Table[IdentityMatrix[2], 5], s[[m]], i]   
  Si[i_, m_] := KroneckerProduct @@ Slist[i, m] 
  SS[i_, j_] := Sum[Si[i, m].Si[j, m], {m, 1, 3}] 
  \end{verbatim}
Note that Slist is a list of 6 matrices, with Pauli spin/isospin matrices at the position $i$.
Next like (Si) promotes it to an operator in the ``monom space". In such notations complex operators can be written as the usual sum of their products, e.g. the spin-spin interaction is $(\vec S_1 \vec S_2)\rightarrow SS[1,2]$. )

   Before we present the wave functions for these states, we show
   their ``coordinate density" $| \psi |^2$ summed over all indices,  
\ba | N^*_{J={5 \over 2},J_z=1/2,S=3/2}|^2&\sim &  (2/5) (\lambda_1^2 + \lambda_2^2 + 
   3 \lambda_3^2 + \rho_1^2 + \rho_2^2 +3 \rho_3^2)\nonumber \\
   | N^*_{J=3/2,J_z=1/2,S=3/2}|^2&\sim &  (2/15) (7\lambda_1^2 + 7\lambda_2^2 + 
    \lambda_3^2 + 7\rho_1^2 + 7\rho_2^2 +\rho_3^2)\nonumber \\
| N^*_{J=1/2,J_z=1/2,S=3/2}|^2&\sim &  (2/3) (\lambda_1^2 + \lambda_2^2 + \lambda_3^2 + \rho_1^2 + \rho_2^2 + \rho_3^2) \nonumber \\
| N^*_{J=3/2,J_z=1/2,S=1/2}|^2&\sim &  (1/3) ( \lambda_1^2 +  \lambda_2^2 + 4\lambda_3^2 + 
    \rho_1^2 +  \rho_2^2 +4 \rho_3^2) \nonumber \\
| N^*_{J=1/2,J_z=1/2,S=1/2}|^2&\sim &  (2/3) (\lambda_1^2 + \lambda_2^2 + \lambda_3^2 + \rho_1^2 + \rho_2^2 + \rho_3^2) \nonumber 
\ea 
Note that all the $J=\frac 12$ states have 6d spherical shape, as they should, while the $J=\frac 52$ have deformations of opposite signs to both the $J=\frac 32$ states.

\end{widetext}

  \begin{figure}[h]
\begin{center}
\includegraphics[width=9cm]{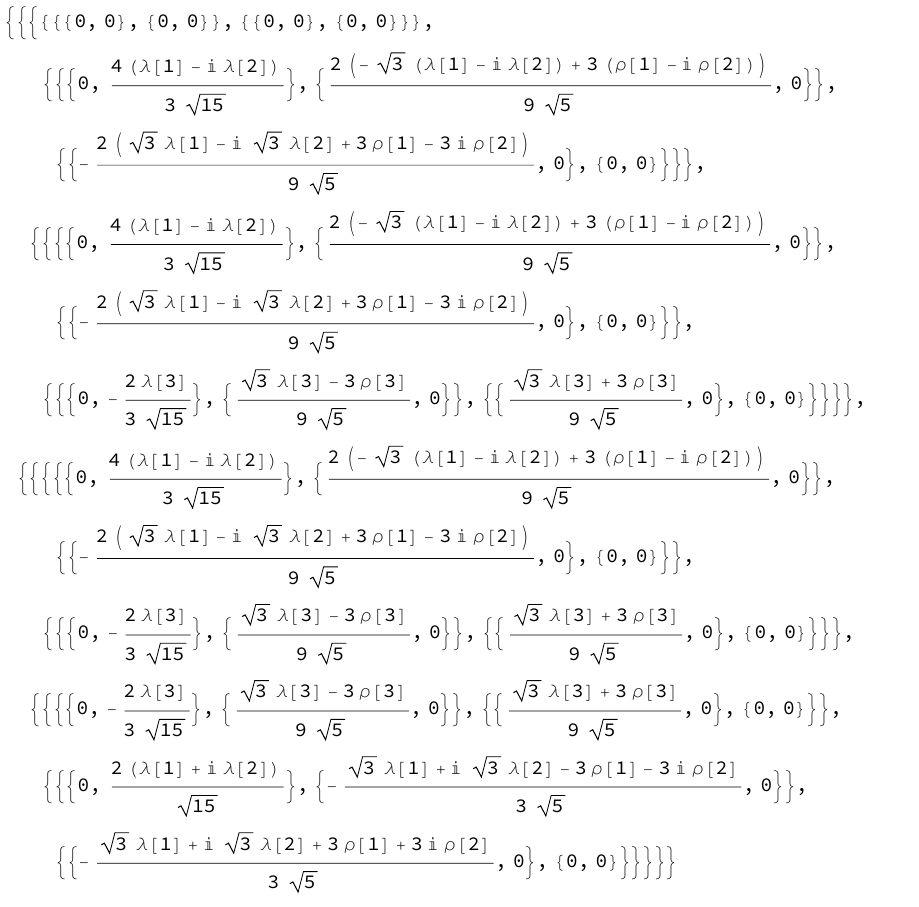}
\caption{ Orbital-Orbital-spin-isospin wave function for \\$ | N^*, J={5 \over 2},Jz=1/2,S=3/2 > $ .}
\label{fig_N51S3}
\end{center}
\end{figure}
  \begin{figure}[h]
\begin{center}
\includegraphics[width=9cm]{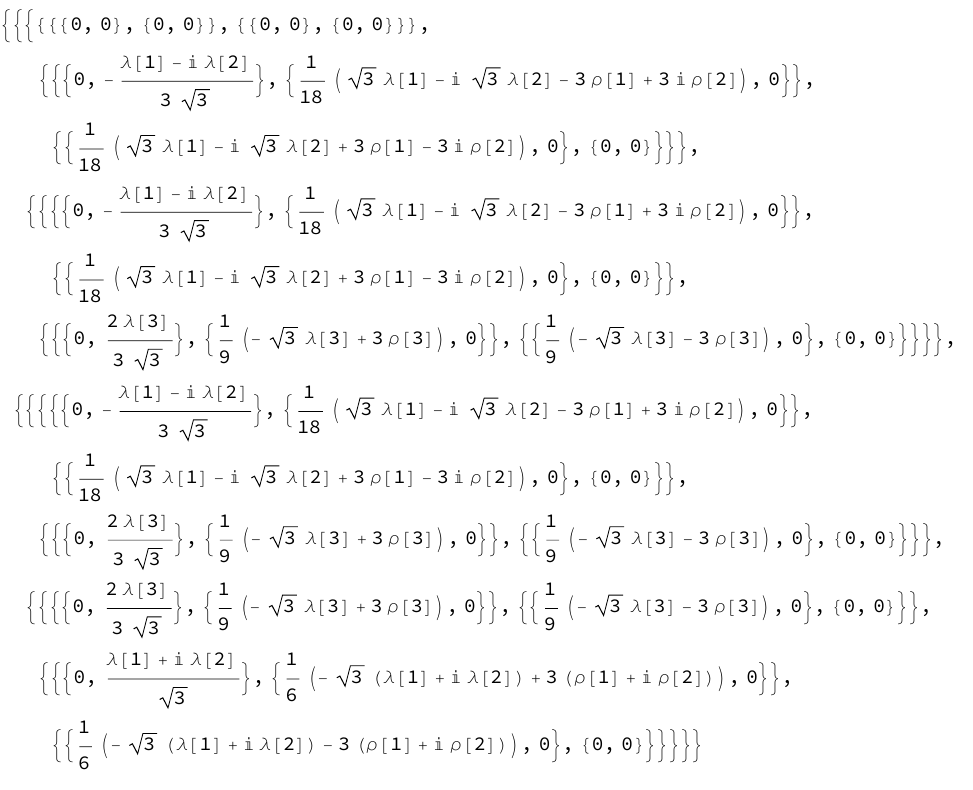}
\caption{ Orbital-spin-isospin wave function for \\$ | N^*, J=3/2,Jz=1/2,S=3/2 > $ .}
\label{fig_N31S3}
\end{center}
\end{figure}
  \begin{figure}[h]
\begin{center}
\includegraphics[width=9cm]{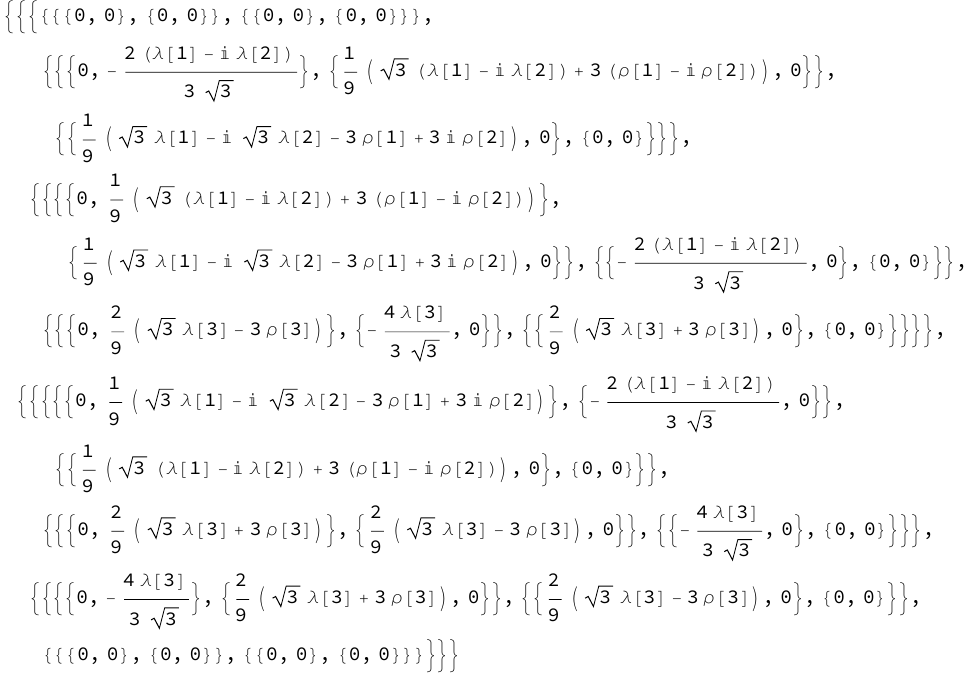}
\caption{ Orbital-spin-isospin wave function for \\$ | N^*, J=1/2,Jz=1/2,S=3/2 > $ .}
\label{fig_N11S3}
\end{center}
\end{figure}
  
  \begin{figure}[h]
\begin{center}
\includegraphics[width=9cm]{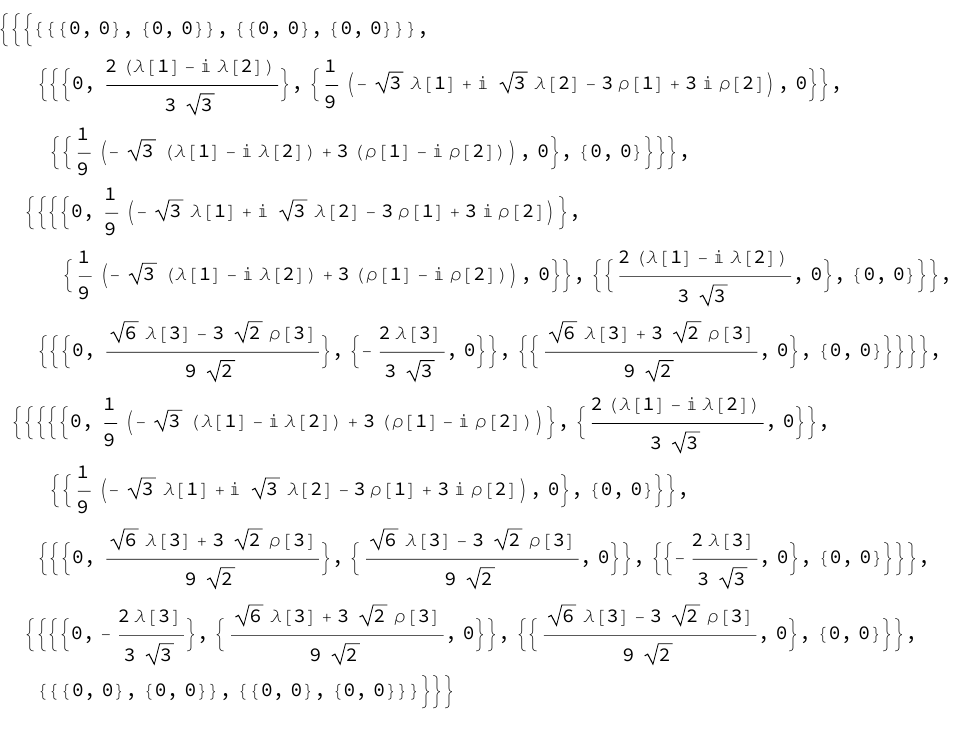}
\caption{ Orbital-spin-isospin wave function for \\$ | N^*, J=3/2,Jz=1/2,S=1/2 > $ .}
\label{fig_N31S1}
\end{center}
\end{figure}

  \begin{figure}[h]
\begin{center}
\includegraphics[width=9cm]{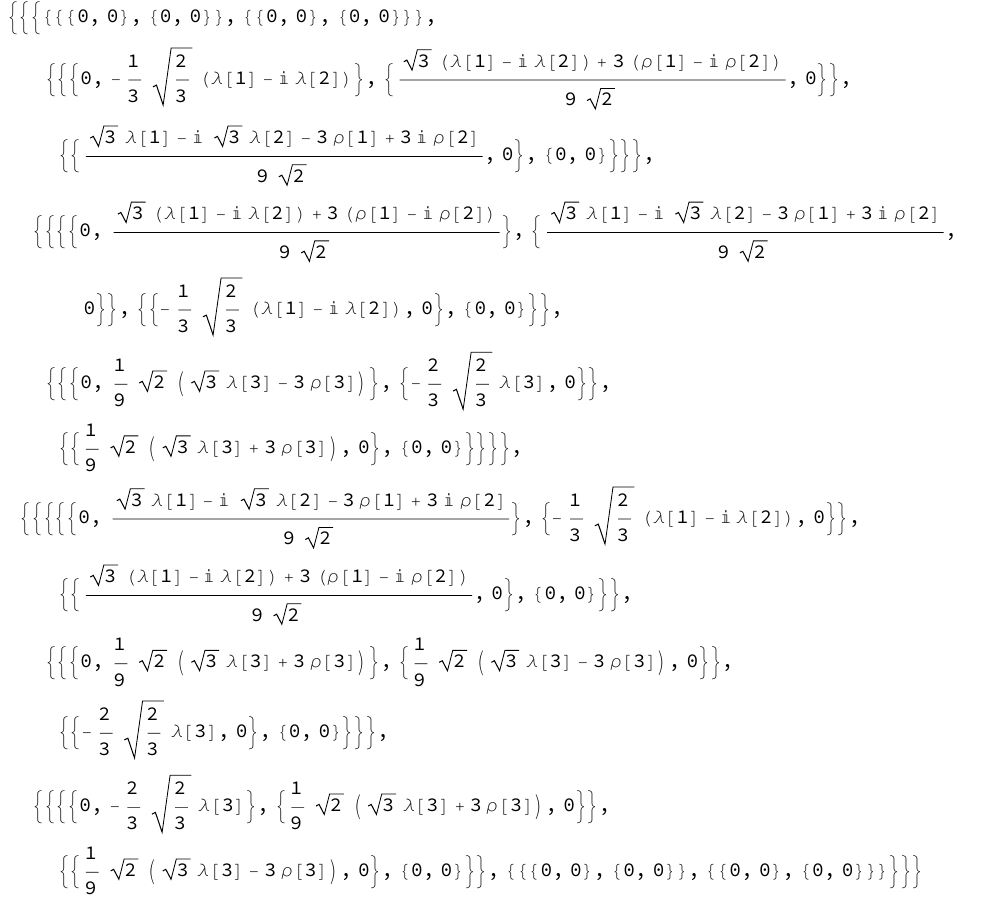}
\caption{Orbital-spin-isospin wave function for \\$ | N^*, J=1/2,Jz=1/2,S=1/2 > $}
\label{fig_N11S1}
\end{center}
\end{figure}

\newpage
\bibliography{spin_baryons}

\begin{thebibliography}{23}%
\makeatletter
\providecommand \@ifxundefined [1]{%
 \@ifx{#1\undefined}
}%
\providecommand \@ifnum [1]{%
 \ifnum #1\expandafter \@firstoftwo
 \else \expandafter \@secondoftwo
 \fi
}%
\providecommand \@ifx [1]{%
 \ifx #1\expandafter \@firstoftwo
 \else \expandafter \@secondoftwo
 \fi
}%
\providecommand \natexlab [1]{#1}%
\providecommand \enquote  [1]{``#1''}%
\providecommand \bibnamefont  [1]{#1}%
\providecommand \bibfnamefont [1]{#1}%
\providecommand \citenamefont [1]{#1}%
\providecommand \href@noop [0]{\@secondoftwo}%
\providecommand \href [0]{\begingroup \@sanitize@url \@href}%
\providecommand \@href[1]{\@@startlink{#1}\@@href}%
\providecommand \@@href[1]{\endgroup#1\@@endlink}%
\providecommand \@sanitize@url [0]{\catcode `\\12\catcode `\$12\catcode
  `\&12\catcode `\#12\catcode `\^12\catcode `\_12\catcode `\%12\relax}%
\providecommand \@@startlink[1]{}%
\providecommand \@@endlink[0]{}%
\providecommand \url  [0]{\begingroup\@sanitize@url \@url }%
\providecommand \@url [1]{\endgroup\@href {#1}{\urlprefix }}%
\providecommand \urlprefix  [0]{URL }%
\providecommand \Eprint [0]{\href }%
\providecommand \doibase [0]{http://dx.doi.org/}%
\providecommand \selectlanguage [0]{\@gobble}%
\providecommand \bibinfo  [0]{\@secondoftwo}%
\providecommand \bibfield  [0]{\@secondoftwo}%
\providecommand \translation [1]{[#1]}%
\providecommand \BibitemOpen [0]{}%
\providecommand \bibitemStop [0]{}%
\providecommand \bibitemNoStop [0]{.\EOS\space}%
\providecommand \EOS [0]{\spacefactor3000\relax}%
\providecommand \BibitemShut  [1]{\csname bibitem#1\endcsname}%
\let\auto@bib@innerbib\@empty
%</preamble>
\bibitem [{\citenamefont {Isgur}\ and\ \citenamefont
  {Karl}(1979{\natexlab{a}})}]{Isgur:1978wd}%
  \BibitemOpen
  \bibfield  {author} {\bibinfo {author} {\bibfnamefont {N.}~\bibnamefont
  {Isgur}}\ and\ \bibinfo {author} {\bibfnamefont {G.}~\bibnamefont {Karl}},\
  }\href {\doibase 10.1103/PhysRevD.19.2653} {\bibfield  {journal} {\bibinfo
  {journal} {Phys. Rev. D}\ }\textbf {\bibinfo {volume} {19}},\ \bibinfo
  {pages} {2653} (\bibinfo {year} {1979}{\natexlab{a}})},\ \bibinfo {note}
  {[Erratum: Phys.Rev.D 23, 817 (1981)]}\BibitemShut {NoStop}%
\bibitem [{\citenamefont {Isgur}\ and\ \citenamefont
  {Karl}(1979{\natexlab{b}})}]{Isgur:1979be}%
  \BibitemOpen
  \bibfield  {author} {\bibinfo {author} {\bibfnamefont {N.}~\bibnamefont
  {Isgur}}\ and\ \bibinfo {author} {\bibfnamefont {G.}~\bibnamefont {Karl}},\
  }\href {\doibase 10.1103/PhysRevD.20.1191} {\bibfield  {journal} {\bibinfo
  {journal} {Phys. Rev. D}\ }\textbf {\bibinfo {volume} {20}},\ \bibinfo
  {pages} {1191} (\bibinfo {year} {1979}{\natexlab{b}})}\BibitemShut {NoStop}%
\bibitem [{\citenamefont {Isgur}\ and\ \citenamefont
  {Karl}(1978)}]{Isgur:1978xj}%
  \BibitemOpen
  \bibfield  {author} {\bibinfo {author} {\bibfnamefont {N.}~\bibnamefont
  {Isgur}}\ and\ \bibinfo {author} {\bibfnamefont {G.}~\bibnamefont {Karl}},\
  }\href {\doibase 10.1103/PhysRevD.18.4187} {\bibfield  {journal} {\bibinfo
  {journal} {Phys. Rev. D}\ }\textbf {\bibinfo {volume} {18}},\ \bibinfo
  {pages} {4187} (\bibinfo {year} {1978})}\BibitemShut {NoStop}%
\bibitem [{\citenamefont {Capstick}\ and\ \citenamefont
  {Isgur}(1985)}]{Capstick:1985xss}%
  \BibitemOpen
  \bibfield  {author} {\bibinfo {author} {\bibfnamefont {S.}~\bibnamefont
  {Capstick}}\ and\ \bibinfo {author} {\bibfnamefont {N.}~\bibnamefont
  {Isgur}},\ }\href {\doibase 10.1063/1.35361} {\bibfield  {journal} {\bibinfo
  {journal} {AIP Conf. Proc.}\ }\textbf {\bibinfo {volume} {132}},\ \bibinfo
  {pages} {267} (\bibinfo {year} {1985})}\BibitemShut {NoStop}%
\bibitem [{\citenamefont {Ji}\ \emph {et~al.}(2003)\citenamefont {Ji},
  \citenamefont {Ma},\ and\ \citenamefont {Yuan}}]{Ji:2002xn}%
  \BibitemOpen
  \bibfield  {author} {\bibinfo {author} {\bibfnamefont {X.-d.}\ \bibnamefont
  {Ji}}, \bibinfo {author} {\bibfnamefont {J.-P.}\ \bibnamefont {Ma}}, \ and\
  \bibinfo {author} {\bibfnamefont {F.}~\bibnamefont {Yuan}},\ }\href {\doibase
  10.1016/S0550-3213(03)00010-5} {\bibfield  {journal} {\bibinfo  {journal}
  {Nucl. Phys. B}\ }\textbf {\bibinfo {volume} {652}},\ \bibinfo {pages} {383}
  (\bibinfo {year} {2003})},\ \Eprint {http://arxiv.org/abs/hep-ph/0210430}
  {arXiv:hep-ph/0210430} \BibitemShut {NoStop}%
\bibitem [{\citenamefont {Shuryak}\ and\ \citenamefont
  {Zahed}(2023{\natexlab{a}})}]{Shuryak:2021fsu}%
  \BibitemOpen
  \bibfield  {author} {\bibinfo {author} {\bibfnamefont {E.}~\bibnamefont
  {Shuryak}}\ and\ \bibinfo {author} {\bibfnamefont {I.}~\bibnamefont
  {Zahed}},\ }\href {\doibase 10.1103/PhysRevD.107.034023} {\bibfield
  {journal} {\bibinfo  {journal} {Phys. Rev. D}\ }\textbf {\bibinfo {volume}
  {107}},\ \bibinfo {pages} {034023} (\bibinfo {year} {2023}{\natexlab{a}})},\
  \Eprint {http://arxiv.org/abs/2110.15927} {arXiv:2110.15927 [hep-ph]}
  \BibitemShut {NoStop}%
\bibitem [{\citenamefont {Shuryak}\ and\ \citenamefont
  {Zahed}(2023{\natexlab{b}})}]{Shuryak:2022thi}%
  \BibitemOpen
  \bibfield  {author} {\bibinfo {author} {\bibfnamefont {E.}~\bibnamefont
  {Shuryak}}\ and\ \bibinfo {author} {\bibfnamefont {I.}~\bibnamefont
  {Zahed}},\ }\href {\doibase 10.1103/PhysRevD.107.034026} {\bibfield
  {journal} {\bibinfo  {journal} {Phys. Rev. D}\ }\textbf {\bibinfo {volume}
  {107}},\ \bibinfo {pages} {034026} (\bibinfo {year} {2023}{\natexlab{b}})},\
  \Eprint {http://arxiv.org/abs/2202.00167} {arXiv:2202.00167 [hep-ph]}
  \BibitemShut {NoStop}%
\bibitem [{\citenamefont {Gross}\ \emph {et~al.}(2022)\citenamefont {Gross}
  \emph {et~al.}}]{Gross:2022hyw}%
  \BibitemOpen
  \bibfield  {author} {\bibinfo {author} {\bibfnamefont {F.}~\bibnamefont
  {Gross}} \emph {et~al.},\ }\href@noop {} {\  (\bibinfo {year} {2022})},\
  \Eprint {http://arxiv.org/abs/2212.11107} {arXiv:2212.11107 [hep-ph]}
  \BibitemShut {NoStop}%
\bibitem [{\citenamefont {Shuryak}\ and\ \citenamefont
  {Rosner}(1989)}]{Shuryak:1988bf}%
  \BibitemOpen
  \bibfield  {author} {\bibinfo {author} {\bibfnamefont {E.~V.}\ \bibnamefont
  {Shuryak}}\ and\ \bibinfo {author} {\bibfnamefont {J.~L.}\ \bibnamefont
  {Rosner}},\ }\href {\doibase 10.1016/0370-2693(89)90477-2} {\bibfield
  {journal} {\bibinfo  {journal} {Phys. Lett. B}\ }\textbf {\bibinfo {volume}
  {218}},\ \bibinfo {pages} {72} (\bibinfo {year} {1989})}\BibitemShut
  {NoStop}%
\bibitem [{\citenamefont {Glozman}\ \emph {et~al.}(1998)\citenamefont
  {Glozman}, \citenamefont {Papp}, \citenamefont {Plessas}, \citenamefont
  {Varga},\ and\ \citenamefont {Wagenbrunn}}]{Glozman:1997fs}%
  \BibitemOpen
  \bibfield  {author} {\bibinfo {author} {\bibfnamefont {L.~Y.}\ \bibnamefont
  {Glozman}}, \bibinfo {author} {\bibfnamefont {Z.}~\bibnamefont {Papp}},
  \bibinfo {author} {\bibfnamefont {W.}~\bibnamefont {Plessas}}, \bibinfo
  {author} {\bibfnamefont {K.}~\bibnamefont {Varga}}, \ and\ \bibinfo {author}
  {\bibfnamefont {R.~F.}\ \bibnamefont {Wagenbrunn}},\ }\href {\doibase
  10.1103/PhysRevC.57.3406} {\bibfield  {journal} {\bibinfo  {journal} {Phys.
  Rev. C}\ }\textbf {\bibinfo {volume} {57}},\ \bibinfo {pages} {3406}
  (\bibinfo {year} {1998})},\ \Eprint {http://arxiv.org/abs/nucl-th/9705011}
  {arXiv:nucl-th/9705011} \BibitemShut {NoStop}%
\bibitem [{\citenamefont {Zahed}\ and\ \citenamefont
  {Brown}(1986)}]{Zahed:1986qz}%
  \BibitemOpen
  \bibfield  {author} {\bibinfo {author} {\bibfnamefont {I.}~\bibnamefont
  {Zahed}}\ and\ \bibinfo {author} {\bibfnamefont {G.~E.}\ \bibnamefont
  {Brown}},\ }\href {\doibase 10.1016/0370-1573(86)90142-0} {\bibfield
  {journal} {\bibinfo  {journal} {Phys. Rept.}\ }\textbf {\bibinfo {volume}
  {142}},\ \bibinfo {pages} {1} (\bibinfo {year} {1986})}\BibitemShut {NoStop}%
\bibitem [{\citenamefont {Liu}\ \emph {et~al.}(2021{\natexlab{a}})\citenamefont
  {Liu}, \citenamefont {Nowak},\ and\ \citenamefont {Zahed}}]{Liu:2021tpq}%
  \BibitemOpen
  \bibfield  {author} {\bibinfo {author} {\bibfnamefont {Y.}~\bibnamefont
  {Liu}}, \bibinfo {author} {\bibfnamefont {M.~A.}\ \bibnamefont {Nowak}}, \
  and\ \bibinfo {author} {\bibfnamefont {I.}~\bibnamefont {Zahed}},\ }\href
  {\doibase 10.1103/PhysRevD.104.114021} {\bibfield  {journal} {\bibinfo
  {journal} {Phys. Rev. D}\ }\textbf {\bibinfo {volume} {104}},\ \bibinfo
  {pages} {114021} (\bibinfo {year} {2021}{\natexlab{a}})},\ \Eprint
  {http://arxiv.org/abs/2108.04334} {arXiv:2108.04334 [hep-ph]} \BibitemShut
  {NoStop}%
\bibitem [{\citenamefont {Liu}\ \emph {et~al.}(2021{\natexlab{b}})\citenamefont
  {Liu}, \citenamefont {Nowak},\ and\ \citenamefont {Zahed}}]{Liu:2021ixf}%
  \BibitemOpen
  \bibfield  {author} {\bibinfo {author} {\bibfnamefont {Y.}~\bibnamefont
  {Liu}}, \bibinfo {author} {\bibfnamefont {M.~A.}\ \bibnamefont {Nowak}}, \
  and\ \bibinfo {author} {\bibfnamefont {I.}~\bibnamefont {Zahed}},\ }\href
  {\doibase 10.1103/PhysRevD.104.114022} {\bibfield  {journal} {\bibinfo
  {journal} {Phys. Rev. D}\ }\textbf {\bibinfo {volume} {104}},\ \bibinfo
  {pages} {114022} (\bibinfo {year} {2021}{\natexlab{b}})},\ \Eprint
  {http://arxiv.org/abs/2108.07074} {arXiv:2108.07074 [hep-ph]} \BibitemShut
  {NoStop}%
\bibitem [{\citenamefont {Close}(1979)}]{Close:1979bt}%
  \BibitemOpen
  \bibfield  {author} {\bibinfo {author} {\bibfnamefont {F.~E.}\ \bibnamefont
  {Close}},\ }\href@noop {} {\emph {\bibinfo {title} {{An Introduction to
  Quarks and Partons}}}}\ (\bibinfo {year} {1979})\BibitemShut {NoStop}%
\bibitem [{\citenamefont {Capstick}\ and\ \citenamefont
  {Roberts}(2000)}]{Capstick:2000qj}%
  \BibitemOpen
  \bibfield  {author} {\bibinfo {author} {\bibfnamefont {S.}~\bibnamefont
  {Capstick}}\ and\ \bibinfo {author} {\bibfnamefont {W.}~\bibnamefont
  {Roberts}},\ }\href {\doibase 10.1016/S0146-6410(00)00109-5} {\bibfield
  {journal} {\bibinfo  {journal} {Prog. Part. Nucl. Phys.}\ }\textbf {\bibinfo
  {volume} {45}},\ \bibinfo {pages} {S241} (\bibinfo {year} {2000})},\ \Eprint
  {http://arxiv.org/abs/nucl-th/0008028} {arXiv:nucl-th/0008028} \BibitemShut
  {NoStop}%
\bibitem [{\citenamefont {Miesch}\ \emph {et~al.}(2023)\citenamefont {Miesch},
  \citenamefont {Shuryak},\ and\ \citenamefont {Zahed}}]{Miesch:2023hjt}%
  \BibitemOpen
  \bibfield  {author} {\bibinfo {author} {\bibfnamefont {N.}~\bibnamefont
  {Miesch}}, \bibinfo {author} {\bibfnamefont {E.}~\bibnamefont {Shuryak}}, \
  and\ \bibinfo {author} {\bibfnamefont {I.}~\bibnamefont {Zahed}},\
  }\href@noop {} {\  (\bibinfo {year} {2023})},\ \Eprint
  {http://arxiv.org/abs/2308.05638} {arXiv:2308.05638 [hep-ph]} \BibitemShut
  {NoStop}%
\bibitem [{\citenamefont {Eichten}\ and\ \citenamefont
  {Feinberg}(1981)}]{Eichten:1980mw}%
  \BibitemOpen
  \bibfield  {author} {\bibinfo {author} {\bibfnamefont {E.}~\bibnamefont
  {Eichten}}\ and\ \bibinfo {author} {\bibfnamefont {F.}~\bibnamefont
  {Feinberg}},\ }\href {\doibase 10.1103/PhysRevD.23.2724} {\bibfield
  {journal} {\bibinfo  {journal} {Phys. Rev. D}\ }\textbf {\bibinfo {volume}
  {23}},\ \bibinfo {pages} {2724} (\bibinfo {year} {1981})}\BibitemShut
  {NoStop}%
\bibitem [{\citenamefont {Shuryak}\ and\ \citenamefont
  {Zahed}(2023{\natexlab{c}})}]{Shuryak:2023siq}%
  \BibitemOpen
  \bibfield  {author} {\bibinfo {author} {\bibfnamefont {E.}~\bibnamefont
  {Shuryak}}\ and\ \bibinfo {author} {\bibfnamefont {I.}~\bibnamefont
  {Zahed}},\ }\href {\doibase 10.1103/PhysRevD.107.094005} {\bibfield
  {journal} {\bibinfo  {journal} {Phys. Rev. D}\ }\textbf {\bibinfo {volume}
  {107}},\ \bibinfo {pages} {094005} (\bibinfo {year} {2023}{\natexlab{c}})},\
  \Eprint {http://arxiv.org/abs/2301.12238} {arXiv:2301.12238 [hep-ph]}
  \BibitemShut {NoStop}%
\bibitem [{\citenamefont {Shuryak}\ and\ \citenamefont
  {Zahed}(2023{\natexlab{d}})}]{Shuryak:2022wtk}%
  \BibitemOpen
  \bibfield  {author} {\bibinfo {author} {\bibfnamefont {E.}~\bibnamefont
  {Shuryak}}\ and\ \bibinfo {author} {\bibfnamefont {I.}~\bibnamefont
  {Zahed}},\ }\href {\doibase 10.1103/PhysRevD.107.034027} {\bibfield
  {journal} {\bibinfo  {journal} {Phys. Rev. D}\ }\textbf {\bibinfo {volume}
  {107}},\ \bibinfo {pages} {034027} (\bibinfo {year} {2023}{\natexlab{d}})},\
  \Eprint {http://arxiv.org/abs/2208.04428} {arXiv:2208.04428 [hep-ph]}
  \BibitemShut {NoStop}%
\bibitem [{\citenamefont {Shuryak}\ and\ \citenamefont
  {Zahed}(2023{\natexlab{e}})}]{Shuryak:2021hng}%
  \BibitemOpen
  \bibfield  {author} {\bibinfo {author} {\bibfnamefont {E.}~\bibnamefont
  {Shuryak}}\ and\ \bibinfo {author} {\bibfnamefont {I.}~\bibnamefont
  {Zahed}},\ }\href {\doibase 10.1103/PhysRevD.107.034024} {\bibfield
  {journal} {\bibinfo  {journal} {Phys. Rev. D}\ }\textbf {\bibinfo {volume}
  {107}},\ \bibinfo {pages} {034024} (\bibinfo {year} {2023}{\natexlab{e}})},\
  \Eprint {http://arxiv.org/abs/2111.01775} {arXiv:2111.01775 [hep-ph]}
  \BibitemShut {NoStop}%
\bibitem [{\citenamefont {Shuryak}\ and\ \citenamefont
  {Zahed}(2023{\natexlab{f}})}]{Shuryak:2023fjj}%
  \BibitemOpen
  \bibfield  {author} {\bibinfo {author} {\bibfnamefont {E.}~\bibnamefont
  {Shuryak}}\ and\ \bibinfo {author} {\bibfnamefont {I.}~\bibnamefont
  {Zahed}},\ }\href {\doibase 10.1103/PhysRevD.107.034025} {\bibfield
  {journal} {\bibinfo  {journal} {Phys. Rev. D}\ }\textbf {\bibinfo {volume}
  {107}},\ \bibinfo {pages} {034025} (\bibinfo {year}
  {2023}{\natexlab{f}})}\BibitemShut {NoStop}%
\bibitem [{\citenamefont {Ji}\ \emph {et~al.}(2004)\citenamefont {Ji},
  \citenamefont {Ma},\ and\ \citenamefont {Yuan}}]{Ji:2003yj}%
  \BibitemOpen
  \bibfield  {author} {\bibinfo {author} {\bibfnamefont {X.-d.}\ \bibnamefont
  {Ji}}, \bibinfo {author} {\bibfnamefont {J.-P.}\ \bibnamefont {Ma}}, \ and\
  \bibinfo {author} {\bibfnamefont {F.}~\bibnamefont {Yuan}},\ }\href {\doibase
  10.1140/epjc/s2003-01563-y} {\bibfield  {journal} {\bibinfo  {journal} {Eur.
  Phys. J. C}\ }\textbf {\bibinfo {volume} {33}},\ \bibinfo {pages} {75}
  (\bibinfo {year} {2004})},\ \Eprint {http://arxiv.org/abs/hep-ph/0304107}
  {arXiv:hep-ph/0304107} \BibitemShut {NoStop}%
\bibitem [{\citenamefont {Belitsky}\ and\ \citenamefont
  {Radyushkin}(2005)}]{Belitsky:2005qn}%
  \BibitemOpen
  \bibfield  {author} {\bibinfo {author} {\bibfnamefont {A.~V.}\ \bibnamefont
  {Belitsky}}\ and\ \bibinfo {author} {\bibfnamefont {A.~V.}\ \bibnamefont
  {Radyushkin}},\ }\href {\doibase 10.1016/j.physrep.2005.06.002} {\bibfield
  {journal} {\bibinfo  {journal} {Phys. Rept.}\ }\textbf {\bibinfo {volume}
  {418}},\ \bibinfo {pages} {1} (\bibinfo {year} {2005})},\ \Eprint
  {http://arxiv.org/abs/hep-ph/0504030} {arXiv:hep-ph/0504030} \BibitemShut
  {NoStop}%
\end{thebibliography}%
\end{document}